\documentclass[sigconf,nonacm]{acmart}
\AtBeginDocument{%
  \providecommand\BibTeX{{%
    \normalfont B\kern-0.5em{\scshape i\kern-0.25em b}\kern-0.8em\TeX}}}

\usepackage{array}

\usepackage{hyperref}
\usepackage{url}
\usepackage{threeparttable,booktabs}

\usepackage{tabularx}
\usepackage[font=footnotesize,labelfont=bf]{caption}
\usepackage{subcaption}

\usepackage[boxed,ruled,vlined,linesnumbered]{algorithm2e}
\usepackage{algorithmic}

\usepackage{multirow}
\usepackage{graphicx}

\usepackage{enumitem}
\setlist[itemize]{leftmargin=*, noitemsep, nolistsep, topsep=0pt}
\setlist[enumerate]{leftmargin=*, noitemsep, nolistsep, topsep=0pt}

\newcommand{\new}[1]{{\color{black}{#1}}}

\def\BibTeX{{\rm B\kern-.05em{\sc i\kern-.025em b}\kern-.08em
    T\kern-.1667em\lower.7ex\hbox{E}\kern-.125emX}}

\title{FedSecurity: A Benchmark for Attacks and Defenses in Federated Learning and Federated LLMs}

\author{Shanshan Han}
\affiliation{\institution{University of California, Irvine}
\city{Irvine}
\country{USA}
}
\email{shanshan.han@uci.edu}

\author{Baturalp Buyukates}
\affiliation{\institution{University of Southern California}
\city{Los Angeles}
\country{USA}
}
\email{buyukate@usc.edu}

\author{Zijian Hu}
\affiliation{\institution{TensorOpera Inc.}
\city{Palo Alto}
\country{USA}
}
\email{zjh@tensoropera.com}

\author{Han Jin}
\affiliation{\institution{University of Southern California}
\city{Los Angeles}
\country{USA}
}
\email{hanjin@usc.edu}

\author{Weizhao Jin}
\affiliation{\institution{University of Southern California}
\city{Los Angeles}
\country{USA}}
\email{weizhaoj@usc.edu}

\author{Lichao Sun}
\affiliation{\institution{Lehigh University}
\city{Bethlehem}
\country{USA}}
\email{lis221@lehigh.edu}

\author{Xiaoyang Wang}
\affiliation{\institution{UIUC}
\city{Urbana}
\country{USA}}
\email{xw28@illinois.edu}

\author{Wenxuan Wu}
\affiliation{\institution{Texas A\&M University}
\city{College Station}
\country{USA}}
\email{ww6726@tamu.edu}

\author{Chulin Xie}
\affiliation{\institution{UIUC}
\city{Urbana}
\country{USA}}
\email{chulinx2@illinois.edu}

\author{Yuhang Yao}
\affiliation{\institution{Carnegie Mellon University}
\city{Pittsburgh}
\country{USA}}
\email{yuhangya@andrew.cmu.edu}

\author{Kai Zhang}
\affiliation{\institution{Lehigh University}
\city{Bethlehem}
\country{USA}}
\email{kaz321@lehigh.edu}

\author{Qifan Zhang}
\affiliation{\institution{University of California, Irvine}
\city{Irvine}
\country{USA}}
\email{qifan.zhang@uci.edu}

\author{Yuhui Zhang}
\affiliation{\institution{Zhejiang University}
\city{Hangzhou}
\country{China}}
\email{zhangyuhui42@zju.edu.cn}

\author{Carlee Joe-Wong}
\affiliation{\institution{Carnegie Mellon University}
\city{Pittsburgh}
\country{USA}}
\email{cjoewong@andrew.cmu.edu}

\author{Salman Avestimehr}
\additionalaffiliation{%
\institution{TensorOpera Inc}
\city{Palo Alto}
\country{USA}}
\affiliation{\institution{USC}
\city{Los Angeles}
\country{USA}
}
\email{avestime@usc.edu}

\author{Chaoyang He}
\affiliation{\institution{TensorOpera Inc.}
\city{Palo Alto}
\country{USA}
}
\email{ch@tensoropera.com}

\renewcommand{\shortauthors}{Han, et al.}

\begin{document}

\begin{abstract}
\new{This paper introduces FedSecurity, an end-to-end benchmark that serves as a supplementary component of the FedML library for simulating adversarial attacks and corresponding defense mechanisms in Federated Learning (FL). 
FedSecurity eliminates the need for implementing the fundamental FL procedures, \textit{e}.\textit{g}., FL training and data loading, from scratch, thus enables users to focus on developing their own attack and defense strategies.
It contains two key components, including FedAttacker that conducts a variety of attacks during FL training, and FedDefender that implements defensive mechanisms to counteract these attacks. 
FedSecurity has the following features: \textit{i}) 
It offers extensive customization options to accommodate a broad range of machine learning models (\textit{e}.\textit{g}., Logistic Regression, ResNet, and GAN) and FL optimizers (\textit{e}.\textit{g}., FedAVG, FedOPT, and FedNOVA); 
\textit{ii}) it enables exploring the effectiveness of attacks and defenses across different datasets and models; and \textit{iii}) it supports flexible configuration and customization through a configuration file and some APIs. 
We further demonstrate FedSecurity's utility and adaptability through federated training of Large Language Models (LLMs) to showcase its potential on a wide range of complex applications.}

\end{abstract}

\maketitle
\section{Introduction}

Federated Learning (FL)~\citep{mcmahan2017communication} facilitates training across distributed data and enable clients to utilize their local data to train machine learning models collaboratively. Instead of collecting data to a centralized server, FL clients train models on their local data and share the local models with the FL server, where the local models are aggregated into a global model.
FL has attracted considerable attention across various domains and has been utilized in numerous areas such as next-word prediction~\citep{hard2018federated,chen2019federated,ramaswamy2019federated}, hot-word detection~\citep{leroy2019federated}, financial risk assessment~\citep{byrd2020differentially}, and cancer risk prediction~\citep{chowdhury2022review}, etc.

\begin{table*}[htbp]
\centering
\caption{Types of Defenses and Their Implementations Against Specific Attacks}
\label{tab:defenses}
\begin{tabular}{cll}
\toprule
\textbf{Type of Defenses} & \textbf{Implementations} & \textbf{Attacks Against (Selected)} \\
\midrule
\multirow{8}{8em}{Before-aggregation defenses} & 
SLSGD~\citep{xie2020slsgd} & 
Data poisoning attacks, \textit{e}.\textit{g}.,~\citep{label_flipping}, \\ \cmidrule{2-3}
 & Residual Reweighting Defense~\citep{fu2019attack} & Backdoor attacks, \textit{e}.\textit{g}.,~\citep{gu2019badnets, label_flipping} \\\cmidrule{2-3}
 & Foolsgold~\citep{foolsgold} & Backdoor attacks, \textit{e}.\textit{g}.~\citep{label_flipping} \\\cmidrule{2-3}
 & Krum and $m$-Krum~\citep{krum} \quad Bulyan~\citep{bulyan_guerraoui2018hidden}& \multirow{5}{26em}{Model poisoning attacks, \textit{e}.\textit{g}., Byzantine attacks~\citep{chen2017distributed,fang2020local,lin2019free} or Backdoor attacks that attack by poisoning model updates ~\citep{how_to_backdoor}}\\
 & Norm Clipping~\citep{sun2019can}\quad Fl-wbc~\citep{sun2021fl}  &  \\
 &  coordinate-wise median~\citep{yin2018byzantine} \quad CClip~\citep{karimireddy2020byzantine} &  \\
 & coordinate-wise trimmed mean~\citep{yin2018byzantine} &  \\
 & anomaly detection~\citep{anomaly_detection} \quad weak DP~\citep{sun2019can} & \\
 \hline
\addlinespace
\multirow{4}{8em}{On-aggregation defenses } & 
Robust Learning Rate~\citep{ozdayi2021defending} & Backdoor attacks, \textit{e}.\textit{g}.,~\citep{label_flipping} \\ \cmidrule{2-3}
 & SLSGD~\citep{xie2020slsgd} & Data poisoning attacks, \textit{e}.\textit{g}.,~\citep{label_flipping}, \\\cmidrule{2-3}
 & Geometric median~\citep{chen2017distributed}, RFA~\citep{rfa} & Byzantine attacks~\citep{chen2017distributed,fang2020local,lin2019free}, data poisoning attacks~\citep{label_flipping} \\
 \hline
\addlinespace
\multirow{2}{8em}{After-aggregation defenses} & 
CClip~\citep{karimireddy2020byzantine} & Byzantine attacks or backdoor attacks \\\cmidrule{2-3}
 & CRFL~\citep{xie2021crfl} & Backdoor attacks \\
\bottomrule
\end{tabular}
\end{table*}
\begin{table}[ht]
\centering
\captionsetup{justification=centering}
\caption{Attacks Implemented in FedSecurity}
\label{tab:attacks}
\begin{tabular}{p{2.5cm}l}
\toprule
\textbf{Type of Attacks} & \textbf{Implementations} \\
\midrule
\multirow{5}{2.5cm}{Model poisoning} & Byzantine attack~\cite{chen2017distributed,fang2020local,lin2019free}: zero mode, \\ & random mode, and flipping mode \\ 
& Minimizing Distance Backdoor~\cite{baruch2019little} \\ 
& Model Replacement Backdoor~\cite{how_to_backdoor} \\ 
& Lazy Worker (or Free Rider)~\cite{free_rider,fraboni2021free} \\\hline
\multirow{2}{2.6cm}{Data poisoning} & Label Flipping Backdoor attack~\cite{label_flipping} \\
& Edge Case Backdoor Attack~\cite{wang2020attack} \\\hline

\multirow{3}{*}{Data reconstruction} & Deep Leakage Attack~\cite{deep_leakage} \\
& Inverting Gradient Attack~\cite{geiping2020inverting} \\
& Revealing Labels Attack~\cite{dang2021revealing} \\
\bottomrule
\end{tabular}
\end{table}

Recently, FL applications are expanded with large language models (LLMs), \textit{i}.\textit{e}., \emph{federated LLMs}~\citep{chen_fedLLM}. 
Currently, there are industry products that utilize FL (or distributed training) to train LLMs, such as Deepspeed ZeRO~\citep{deepspeed_zero,deepspeed_zero2}, HuggingFace Accelerate~\citep{huggingface_accelrate}, Pytorch Lightning Fabric~\citep{pytorch_lightning}. 
FL facilitates LLM training due to the following reasons: \textit{i}) \textit{Distributed nature of LLM training data:} LLMs are pre-trained using large amounts of data, which often reside in different locations. Collecting such data to a central server is expensive and may also leak sensitive user information, while a viable way is to train LLMs in a federated manner.
\textit{ii}) \textit{Scalability and efficiency:} LLMs, such as GPT-3~\citep{brown2020language}, have an extremely large number of parameters. Training LLMs on a single machine is infeasible and inflexible, while FL can be a good choice. 
\textit{iii}) \textit{Continuous improvement with user data:} LLMs can be deployed in a federated manner and local instances of the models can be further finetuned based on the local data, 
which enables the global model to improve over time based on users' data without ever having direct access to the data. This is particularly relevant for privacy-sensitive fields such as healthcare or personal communications.


FL, as well as federated LLMs, \new{aim to} maintain privacy and security of client data by allowing clients to train locally without sharing their data to other parties. 
However, its decentralized and collaborative nature inadvertently introduce privacy and security vulnerabilities. 
\new{
Adversarial clients compromise the integrity of the global model by submitting spurious models to prevent the global model from converging~\citep{chen2017distributed,fang2020local,lin2019free,baruch2019little,how_to_backdoor,free_rider,fraboni2021free} and/or planting backdoors to induce the global model to mis-classify specific samples~\citep{baruch2019little,how_to_backdoor,label_flipping,wang2020attack}. Adversaries can also reconstruct training data from model updates or gradients~\citep{deep_leakage,geiping2020inverting,dang2021revealing,hu2023source,hu2021source}. }
Meanwhile, a wide range of defense mechanisms has emerged to mitigate the impact of these attacks~\citep{Li2022LoMarAL,kumari2023baybfed,ozdayi2021defending,krum,xie2020slsgd,chen2017distributed,sun2019can,karimireddy2020byzantine,yin2018byzantine,rfa,foolsgold,xie2021crfl,Ma2022ShieldFLMM,Kumar2022FedCleanAD,Chen2022FedDefDA,sun2020ldp,lyu2022privacy,zhang2022efficient}. 
Despite the efforts for addressing the vulnerability of FL systems, there still lacks a comprehensive benchmark for comparing approaches under unified sittings. 
Moreover, while existing works have explored effectiveness of attacks and defenses on small-scale models, there remains a significant gap in understanding how these mechanisms perform against LLMs. Given that LLMs possess a large number of parameters and are trained on complex datasets obtained from unregulated sources, the effectiveness of attacks and defenses may be diminished when applied to LLMs.
These motivate an urgent need for a standardized and comprehensive benchmark to evaluate baseline attack and defense strategies in the context of FL and federated LLMs.

This paper introduces FedSecurity\footnote{Code: https://github.com/FedML-AI/FedML/tree/master/python/fedml/core/security}, a benchmark that simulates attacks and defenses in FedML~\citep{he2020fedml}. 
FedSecurity comprises two primary components: FedAttacker and FedDefender. 
FedAttacker simulates attacks in FL to help understand and prepare for potential security risks, while
FedDefender is equipped with \new{the state-of-the-art} defense mechanisms to counteract the attacks injected by FedAttacker. 
We summarize our contributions as follows:



\noindent\textbf{\textit{i}) Enabling benchmarking of several different attacks and defenses in FL}. FedSecurity implements attacks and defenses that are widely considered in the literature. \new{We summarize the defenses and the attacks in Table~\ref{tab:defenses} and Table~\ref{tab:attacks}, respectively.} 

\begin{figure}[h]
     \centering
     \begin{subfigure}[b]{0.23\textwidth}
         \centering
         \includegraphics[width=\textwidth]{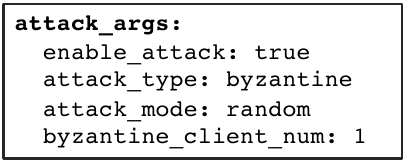}
         \caption{Byzantine attack~\cite{chen2017distributed,fang2020local}.}
         \label{fig:attack_example}
     \end{subfigure}
     \hfill 
     \begin{subfigure}[b]{0.23\textwidth}
         \centering
         \includegraphics[width=\textwidth]{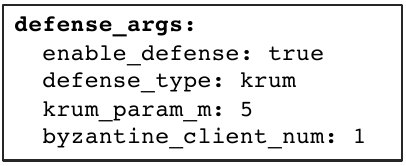}
         \caption{$m$-Krum~\cite{krum}.}
         \label{fig:defense_example}
     \end{subfigure}
     \caption{Examples of attack and defense configurations.}
     \label{fig:examples}
\end{figure}

\noindent\textbf{\textit{ii}) Supporting flexible configuration and customization. }
FedSecurity supports configurations using a .yaml file. 
Sample configurations for attacks and defenses are shown in Figures~\ref{fig:attack_example}
and Figures~\ref{fig:defense_example}, respectively. FedSecurity also provides APIs to enable customizing attacks and defenses. 



\noindent\textbf{\textit{iii}) Supporting various models and FL optimizers.}
FedSecurity can be utilized with a wide range of models, including Logistic Regression, LeNet~\citep{LeNet}, ResNet~\citep{He2015DeepRL}, CNN~\citep{cnn}, RNN~\citep{rnn}, GAN~\citep{Goodfellow2014GenerativeAN}, etc. FedSecurity is compatible with various FL optimizers, such as FedAVG~\citep{McMahan2016CommunicationEfficientLO}, FedSGD~\citep{fedsgd}, FedOPT~\citep{reddi2021adaptive}, FedPROX~\citep{FedProx}, FedGKT~\citep{he2020group}, FedGAN~\citep{rasouli2020fedgan}, FedNAS~\citep{he2021fednas}, FedNOVA~\citep{Wang2020TacklingTO}, etc.

\noindent\textbf{\textit{iv}) Extensions to federated LLMs and real-world applications.} 
FedSecurity can simulate attacks and defenses during training of federated LLMs. It can also be integrated with real-world FL applications; see \textbf{Exp 7}, where we utilize edge devices from Theta Network~\cite{theta_network} as clients instead of simulating on a single machine. 

\noindent\textit{\underline{\textbf{Key takeaways}}}: 
\textit{i})
While defense mechanisms can help mitigate attacks, it might also bring a loss of accuracy to the aggregation results. 
Therefore, when integrating defenses into FL applications, it's crucial to weigh the benefits against potential drawbacks.
\textit{ii}) Nearly all existing defense mechanisms are impractical in real-world FL applications, as they compromise accuracy even if no attack happened. However, attacks happen infrequently in practice.  
A defense strategy that is practical for real-world systems is in need, where the defense should satisfy: 1) it must detect if attacks have happened and only activate the defense mechanism when attacks are detected; and 2) it must identify malicious clients accurately without harming benign local models. \new{\textit{iii}) Based on our findings, we have developed a novel defense strategy; see~\citep{anomaly_detection} for details. }

\section{Preliminaries and Overview}
This section discusses existing FL security frameworks, then introduces adversarial models, and finally overviews FedSecurity.


\subsection{Existing Benchmark Frameworks}\label{sec:related_work}



Recent years, researchers have proposed multiple benchmarks for FL~\citep{blades, tensorflow2015-whitepaper,ziller2021pysyft,liu2021fate,beutel2020flower,lai2022fedscale,roth2022nvidia,reina2021openfl,silva2020fed,ludwig2020ibm,xie2022federatedscope,dimitriadis2022flute}. 
\new{Among these, only Blades~\citep{blades} and FederatedScope~\citep{xie2022federatedscope} study the implications of adversarial attacks in FL. Blades implements a wide range of attacks, such as~\citep{fang2020local,li2019rsa,baruch2019little,xie2020fall,shejwalkar2021manipulating}, as well as corresponding defenses against those attacks, \textit{e}.\textit{g}.,~\citep{xu2022byzantine,karimireddy2021learning,krum,xu2022byzantine}. While Blades focuses more on model poisoning attacks and data poisoning attacks, it fails to include an important line of work, \textit{i}.\textit{e}., data reconstruction attacks.} 
FederatedScope~\citep{xie2022federatedscope} implements data reconstruction attacks that utilize models or gradients to revert sensitive information, including GAN-based leakage attack~\citep{hitaj2017deep}, Passive Property Inference~\citep{melis2019exploiting}, and DLG attack~\citep{deep_leakage}.  
However, it neglects to address attacks prevalent in the research literature, \textit{e}.\textit{g}., Byzantine attacks~\citep{yin2018byzantine,yang2019byzantine}. It also does not include any defense mechanisms for FL. It is worth noting that, while FederatedScope integrates secret-sharing~\citep{beimel2011secret} for enhancing data privacy, it is in the scope of federated analytics~\citep{elkordy2023federated,ramage_2020,fa1,fa2}, instead of FL.

FedSecurity implements attacks that are widely considered in the literature \new{and covers attacks that are injected at different stages of FL, including model poisoning attacks, data poisoning attacks, and data reconstruction attacks}~\citep{chen2017distributed,fang2020local,lin2019free,baruch2019little,how_to_backdoor,free_rider,fraboni2021free,label_flipping,wang2020attack,deep_leakage,geiping2020inverting,dang2021revealing}. It also integrates a wide range of defense mechanisms to protect against the attacks~\citep{sun2019can,ozdayi2021defending,krum,xie2020slsgd,chen2017distributed,sun2019can,karimireddy2020byzantine,yin2018byzantine,rfa,foolsgold,xie2021crfl,yin2018byzantine,anomaly_detection}.
Moreover, FedSecurity offers flexible configurations and APIs, which enables users to customize their attack and defense strategies efficiently.


\subsection{Adversarial Model}\label{sec:adv_model}

Adversaries in FL have two categories, including active adversaries and passive adversaries, corresponding to security risks and privacy threat in FL, respectively.

\noindent\textbf{Active Adversaries.} Active adversaries intentionally manipulate training data or trained models to achieve malicious goals. This might involve altering models to prevent global model from converging (\textit{e}.\textit{g}., Byzantine attacks~\citep{chen2017distributed,fang2020local}), or subtly misclassifying a specific set of samples to minimally impact the overall performance of the global model (\textit{e}.\textit{g}., backdoor attacks~\citep{how_to_backdoor,wang2020attack,zhang2022neurotoxin}). Active adversaries can take different forms, including:
1) malicious clients who manipulate their local models~\citep{how_to_backdoor,chen2017distributed,fang2020local,zhang2022neurotoxin} or submit contrived models without actual training~\citep{free_rider};
2) a global ``sybil''~\citep{label_flipping,foolsgold} that has full access to the FL system and possesses complete knowledge of the entire system, including local models and global models, as well as clients' local datasets. This ``sybil'' may also modify clients' local datasets and their local models submitted to the server; and 
3) external adversaries or hackers that monitor the communication channel between the clients and the server, thereby intercepting and altering local models during the transfer process.



\begin{figure*}
    \centering
    \includegraphics[width=0.96\textwidth]{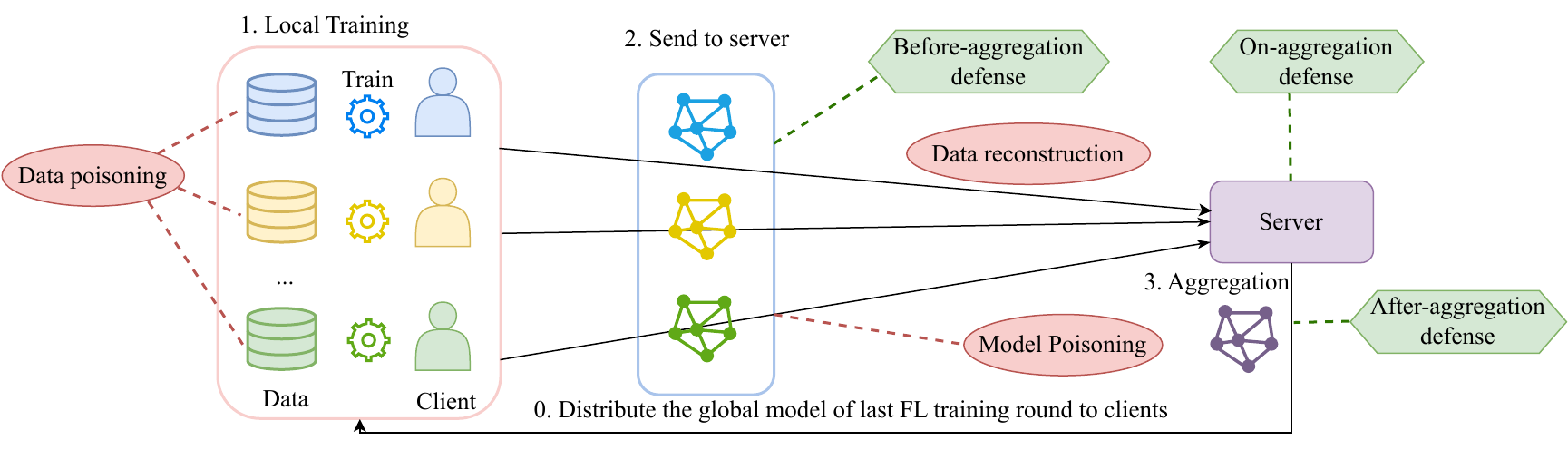}
    \caption{FedSecurity overview. FedSecurity enables injecting attacks/defenses (shown in red/green) at various stages of FL at the clients and at the server.
    }
    \label{fig:overview}
\end{figure*}

\LinesNotNumbered \begin{algorithm}[h]
\textbf{Inputs:} $\mathbf{w}_{g}^{\prime}$: the global model of last FL training round; \\ $\mathcal{W}_l$: local models of the current FL round.

{\bf Variables:}
$\mathcal{A}$: A FedAttacker instance; $\mathcal{D}$: 
A FedDefender instance.

{\bf Function $\mathit{server}\_\mathit{aggregation}(\mathcal{W}_l, \mathbf{w}_g^{\prime})$} \nllabel{ln:server_aggregation_function}
\Begin{

\nl $\mathcal{W}_l\leftarrow \mathit{before}\_\mathit{aggregation}\_\mathit{process}(\mathcal{W}_l, \mathbf{w}_g^{\prime})$

\nl $\mathbf{w}_g\leftarrow \mathit{on}\_\mathit{aggregation}\_\mathit{process}(\mathcal{W}_l, \mathbf{w}_g^{\prime})$

\nl \textbf{return} $\mathit{after}\_\mathit{aggregation}\_\mathit{process}(\mathbf{w}_g)$

}





{\textbf{Function $\mathit{before}\_\mathit{aggregation}\_\mathit{process}(\mathcal{W}_l, \mathbf{w}_g^{\prime})$} \nllabel{ln:before_aggregation}}
\Begin{

\nl \If{$\mathcal{A}$.$\mathit{is}\_\mathit{attack}\_\mathit{enabled}()$}{

\nl \If{$\mathcal{A}$.$\mathit{is}\_\mathit{data}\_\mathit{reconstruction}\_\mathit{attack}()$}{
$\mathcal{A}$.$\mathit{reconstruct}\_\mathit{data}(\mathcal{W}_l, \mathbf{w}_g^{\prime})$
}

\If{$\mathcal{A}$.$\mathit{is}\_\mathit{model}\_\mathit{poisoning}\_\mathit{attack}()$}{$\mathcal{W}_l\leftarrow\mathcal{A}$.$\mathit{poison}\_\mathit{model}(\mathcal{W}_l, \mathbf{w}_g^{\prime})$}

}

\nl \If{$\mathcal{D}$.$\mathit{is}\_\mathit{defense}\_\mathit{enabled}()$ \& $\mathcal{D}$.$\mathit{is}\_\mathit{defense}\_\mathit{before}\_\mathit{aggregation}()$}{$\mathcal{W}_l\leftarrow\mathcal{D}$.$\mathit{defend}\_\mathit{before}\_\mathit{aggregation}(\mathcal{W}_l, \mathbf{w}_g^{\prime})$}

\nl \textbf{return} $\mathcal{W}_l$
}

\textbf{Function $\mathit{on}\_\mathit{aggregation}\_\mathit{process}(\mathcal{W}_l, \mathbf{w}_g)$} \nllabel{ln:on_aggregation}
\Begin{

\nl \If{$\mathcal{D}$.$\mathit{is}\_\mathit{defense}\_\mathit{enabled}()$ \& $\mathcal{D}$.$\mathit{is}\_\mathit{defense}\_\mathit{on}\_\mathit{aggregation}()$}{\textbf{return} $\mathcal{D}$.$\mathit{defend}\_\mathit{on}\_\mathit{aggregation}(\mathcal{W}_l, \mathbf{w}_g)$}

\nl \textbf{return} $\mathit{aggregate}(\mathcal{W}_l)$
}

\textbf{Function $\mathit{after}\_\mathit{aggregation}\_\mathit{process}(\mathbf{w}_g)$} \nllabel{ln:after_aggregation}
\Begin{

\nl \If{$\mathcal{D}$.$\mathit{is}\_\mathit{defense}\_\mathit{enabled}()$ \& $\mathcal{D}$.$\mathit{is}\_\mathit{defense}\_\mathit{after}\_\mathit{aggregation}()$}{\textbf{return} $\mathcal{D}$.$\mathit{defend}\_\mathit{after}\_\mathit{aggregation}(\mathbf{w}_g)$}

\nl \textbf{return} $\mathbf{w}_g$
}

\caption{\textbf{Server Aggregation}}
\label{alg:server_aggregation}
\end{algorithm}

\noindent\textbf{Passive Adversaries.} Passive adversaries do not modify data or models, but can still breach data privacy by inferring sensitive information, such as local training data, from model updates or gradients~\citep{deep_leakage}. Examples of passive adversaries include:
1) an adversarial FL server that tries to guess local training data using submitted local model updates and/or global models;
2) adversarial FL clients that attempt to deduce other clients' training data using the global models provided by the server of each FL training round; and
3) external adversaries, \textit{e}.\textit{g}., hackers, that access communication channels to obtain local and global models transferred between clients and the FL server.

\LinesNotNumbered \begin{algorithm}[h]
\textbf{Inputs:} $\mathit{dataset}$: the local dataset of a client.

{\bf Variables:}
$\mathcal{A}$: A FedAttacker instance.

\nl \textbf{Function $\mathit{client}\_\mathit{training}(\mathit{dataset})$} \nllabel{ln:client_training}
\Begin{

\nl \lIf{$\mathcal{A}$.$\mathit{is}\_\mathit{attack}\_\mathit{enabled}()$ \& $\mathcal{A}$.$\mathit{is}\_\mathit{data}\_\mathit{poisoning}\_\mathit{attack}()$}{$\mathit{dataset}\leftarrow\mathcal{A}$.$\mathit{poison}\_\mathit{data}(\mathit{dataset})$}

\nl $\mathbf{w}_l\leftarrow\mathit{train}(\mathit{dataset})$ 

\nl $\mathit{send}\_\mathit{to}\_\mathit{server}(\mathbf{w}_l)$
}

\caption{\textbf{Client Training}}
\label{alg:local_training}
\end{algorithm}

\new{Adversaries can inject attacks at different stages of FL. Active adversaries can conduct 
\textit{model poisoning attacks} to manipulate local models or \textit{data poisoning attacks} to tamper with local data, while passive adversaries pose privacy threats by
exploiting model updates or gradients, \textit{i}.\textit{e}., \textit{data reconstruction attacks}.} 

\begin{table*}[ht]
\centering
\caption{APIs for Different Types of Attacks in FedSecurity}
\label{tab:attack_api}
\begin{threeparttable}
\begin{tabular}{m{2.8cm}m{5.5cm}m{8cm}}
\toprule
\textbf{Type of Attacks} & \textbf{APIs} & \textbf{Explanations} \\
\midrule

\multirow{3}{8em}{Model Poisoning} &
$\mathit{poison}\_\mathit{model}(\mathit{local}\_\mathit{models}, \mathit{auxiliary}\_\mathit{info})$ &
Take the local models uploaded by clients in the current FL iteration and modify the local models. \\ 
\cmidrule{2-3}

& $\mathit{is}\_\mathit{model}\_\mathit{poisoning}\_\mathit{attack}()$ &
Examine whether FedAttacker is activated and whether it modifies local models. \\

\hline
\multirow{3}{8em}{Data Poisoning} &
$\mathit{poison}\_\mathit{data}(\mathit{dataset})$ &
Take a local dataset and mislabel a set of chosen samples based on the configuration of FedAttacker. \\ 
\cmidrule{2-3}

& $\mathit{is}\_\mathit{data}\_\mathit{poisoning}\_\mathit{attack}()$ &
Examine whether FedAttacker is enabled and whether the attack requires poisoning datasets. \\

\hline
\multirow{3}{9em}{Data Reconstruction} &
$\mathit{reconstruct}\_\mathit{data}(\mathit{model}, \mathit{auxiliary}\_\mathit{info})$ &
Take a client model or a global model, or a model update to reconstruct the training data. \\ 
\cmidrule{2-3}

& $\mathit{is}\_\mathit{data}\_\mathit{reconstruction}\_\mathit{attack}()$ &
Examine whether FedAttacker is enabled and whether it reconstructs training data. \\
\hline
\end{tabular}
\begin{tablenotes}
\small
\item The input \textit{local\_models} is a list of tuples that contain the number of samples and the local models submitted by clients in each FL iteration. 
\item The input $\mathit{auxiliary}\_\mathit{info}$ can be any information that is utilized in the attack functions, \textit{e}.\textit{g}., the global model in the last FL iteration. 
\end{tablenotes}
\end{threeparttable}
\end{table*}





\subsection{Overview of FedSecurity}
FedSecurity serves as an external component of FedML~\citep{he2020fedml} and injects attacks and defenses at different stages of FL training,  without altering the existing FL procedures. FedSecurity utilizes FedAttacker and FedDefender to initiate two instances to simulate attacks and defenses, respectively. Such two instances are initialized once and are accessed by other objects in the FL system, a design achieved using the singleton design pattern~\citep{gamma1995design}.

We summarize the injections of attacks and defenses to the FL framework in FedSecurity in Figure~\ref{fig:overview}. We also provide detailed algorithms for injecting attacks and defenses to different stages of FL training, as shown in Algorithm~\ref{alg:server_aggregation} (for server aggregation) and Algorithm~\ref{alg:local_training} (for client training). Below we introduce injections of attacks
and defenses, respectively. 
\subsubsection{Injection of attacks.} 

Without loss of generality, we classify the attacks into the following categories based on their targets: 

\begin{enumerate}
    \item \textit{Data poisoning attacks} conducted by active adversaries to modify clients' local datasets and are injected at clients~\citep{label_flipping,dang2021revealing}. 
    \item \textit{Model poisoning attacks} 
that are also conducted by active adversaries to temper with local models submitted by clients that participate in the current FL iteration~\citep{fang2020local,shejwalkar2021manipulating,bhagoji2019analyzing}. 
\item \textit{Data reconstruction attacks} 
that are conducted by passive adversaries by exploring local models, model updates, or gradients to infer information about the training data~\citep{Melis2018ExploitingUF, zhang2020gan, luo2021feature,wang2022poisoning,fowl2021robbing}. 

\end{enumerate}

Without loss of generality, FedAttacker injects data poisoning attacks and model poisoning attacks before the aggregation of local models in each FL training round at the server, such that the FedAttacker instance can get access to all client models submitted in the FL training round.
FedAttacker injects data reconstruction attacks at the FL server as well, where the FL server has access to all local models and the global model of each iteration and can perform the attacks with high flexibility. 

\subsubsection{Injection of defenses.} 
FedDefender incorporates defenses to mitigate, if not completely nullify, the impacts of injected attacks. The defenses are designed to address issues related to tampered local models
or manipulated local datasets, which lead to compromised model integrity, 
or information leakage during the exchange of model updates between clients and the FL server.  The defenses manipulate local models and the aggregation procedure to counteract the attacks. 

To facilitate this, FedDefender deploys defenses at the FL server, and provides flexible APIs that enable obtaining all local models and the global model of each FL round while allowing for a customized aggregation process. 
FedDefender utilize three functions at different stages of FL aggregation:

\begin{enumerate}
    \item \textit{Before-aggregation functions} that modify local models at the server before aggregating them.
    \item \textit{On-aggregation functions} that modify the FL aggregation function to mitigate the impacts of malicious local models. 
    \item \textit{After-aggregation functions} that modify the aggregated global model (\textit{e}.\textit{g}., by adding noise or clipping) to protect the real global model or improve its quality.
\end{enumerate}


\section{Implementation of Attacks}

FedAttacker injects model poisoning, data poisoning, and data reconstruction attacks at different stages of FL training and provides APIs for these attacks. We summarize the customization APIs in Table~\ref{tab:attack_api} and introduce each type of attacks in details.

\subsection{Model Poisoning Attacks}


Model poisoning attacks modify the local models uploaded by clients in FL iterations. FedAttacker injects such attacks before FL aggregation in each FL iteration and modifies the local models directly. 
As an example, 
FedAttacker implements three modes of Byzantine attacks~\citep{yin2018byzantine,yang2019byzantine,lin2019free,xu2022byzantine}, as follows. 

\begin{enumerate}
    \item {Zero mode}~\citep{lin2019free} that poisons the client models by setting their weights to zero; 
    \item \textit{Random mode}~\citep{lin2019free} that manipulates client models by attributing random values to model weights; and
    \item  \textit{Flipping mode}~\citep{xu2022byzantine} that updates the global model in the opposite direction by formulating the  local model as $\textbf{w}_g+(\textbf{w}_g-\textbf{w}_{\ell})$, where $\textbf{w}_g$ is the global model, and $\textbf{w}_{\ell}$ is the real local model.
\end{enumerate}

\subsection{Data Poisoning Attacks}
Data poisoning attacks modify local datasets of one or multiple clients to achieve some malicious goals, \textit{e}.\textit{g}., degrading the performance of the global model or inducing the global model to misclassify some samples. As an example, in label flipping attack~\citep{label_flipping}, a global ``sybil'' controls some clients and modifies their local data by mislabeling samples of some classes to wrong classes. Given a source class (or label) $c_\mathit{s}$ and a target class $c_\mathit{t}$, all samples with class $c_\mathit{s}$ on the poisoned clients are re-labeled to an incorrect label $c_\mathit{t}$.

While poisoning local data can be performed by either a global ``sybil'' or  malicious clients, to address a more general case, FedAttacker offers APIs to enable control over each local dataset. 

\begin{table*}[ht]
\centering
\caption{APIs for Different Types of Defenses in FedSecurity}
\label{tab:defense_api}
\begin{threeparttable}
\begin{tabular}{m{2.5cm}m{5.2cm}m{8.5cm}}
\toprule
\textbf{Type of Defenses} & \textbf{APIs} & \textbf{Explanations} \\
\midrule
\multirow{2}{9em}{Before-Aggregation Defenses} &
$\mathit{defend}\_\mathit{before}\_\mathit{aggregation}(\mathit{local}\_\mathit{models}$, $\mathit{auxiliary}\_\mathit{info})$ &
Modify the client models of the current FL iteration to mitigate (or eliminate) the impact of malicious local models. \\\cmidrule{2-3}
& $\mathit{is}\_\mathit{defense}\_\mathit{before}\_\mathit{aggregation}()$ &
Examine whether FedDefender is activated and whether the defense requires injecting functions before aggregating local models at the server. \\
\hline
\multirow{2}{10em}{On-Aggregation Defenses} &
$\mathit{defend}\_\mathit{on}\_\mathit{aggregation}(\mathit{local}\_\mathit{models}$, $\mathit{auxiliary}\_\mathit{info})$ &
Take the local models of the current training round for aggregation and mitigate the impact of malicious clients by modifying aggregation, \textit{e}.\textit{g}., altering aggregation functions. \\\cmidrule{2-3}

& $\mathit{is}\_\mathit{defense}\_\mathit{on}\_\mathit{aggregation}()$ &
Examine if the defense component is enabled and whether the current defense requires the injection of functions during aggregation. \\
\hline
\multirow{2}{10em}{After-Aggregation Defenses} &
$\mathit{defend}\_\mathit{after}\_\mathit{aggregation}(\mathit{global}\_\mathit{model})$ &
Directly modify the global model after aggregation using methods such as clipping or adding noise. \\\cmidrule{2-3}

& $\mathit{is}\_\mathit{defense}\_\mathit{after}\_\mathit{aggregation}()$ &
Examine if the defense component is activated and whether the current defense requires injecting functions after aggregation. \\
\hline
\end{tabular}
\begin{tablenotes}
\small
\item The input \textit{local\_models} is a list of tuples that contain the number of samples and the local models submitted by clients in each FL iteration. 
\item The input $\mathit{auxiliary}\_\mathit{info}$ can be any information that is utilized in the defense functions. 
\end{tablenotes}
\end{threeparttable}
\end{table*}

\subsection{Data Reconstruction Attacks}

Data reconstruction attacks are performed by passive adversaries that attempts to infer sensitive information without actively interfering with the FL training or the local data. 
We assume that there is no leakage during the local training process in FL, as clients are assumed to be on their fully trusted local machines, according to the motivation of FL. As a result, data reconstruction attacks take the trained models (either the global model or the local models) or model updates to revert training data. For example, Deep Leakage from Gradients (DLG)  attack~\citep{deep_leakage} infers local training data from the publicly shared gradients. A passive adversary can use the global model from the previous FL training round and the newly obtained model to compute a ``model update'' between models in different FL training rounds to deduce the training data. 

\subsection{Integration of a New Attack}\label{sec:inject_new_attack}

To customize a new attack, users should follow these steps: \textit{i}) determine the type of the attack, \textit{i}.\textit{e}., model poisoning, data poisoning, or data reconstruction; \textit{ii}) create a new class for the attack 
and implement functions using the APIs, \textit{e}.\textit{g}., $\mathit{attack}\_\mathit{model}(*)$, $\mathit{poison}\_\mathit{data}(*)$, 
and $\mathit{reconstruct}\_\mathit{data}(*)$, to inject attacks at the appropriate stages of FL training; and \textit{iii}) add the attack name to the corresponding enabler functions, \textit{e}.\textit{g}., $\mathit{is}\_\mathit{model}\_\mathit{poisoning}\_\mathit{attack}()$, 
within the FedAttacker class to ensure that the injected attacks are activated at the proper stages of FL training.

\section{Implementation of Defenses}

FedDefender injects defense functions at different stages of FL aggregation at the server. Based on the point of injection, FedDefender provides three types of functions to support defense mechanisms, including 1) before-aggregation, 2) on-aggregation, and 3) after-aggregation. Note that a defense may inject functions at one or multiple stages of FL aggregation. The APIs for defense functions in FedDefender are summarized in Table~\ref{tab:defense_api}.

\subsection{Before-Aggregation Defenses}

Before-aggregation functions operate on local models \new{at the FL server} before aggregating the local models. 
We use Krum~\citep{krum} as an example \new{to demonstrate before-aggregation defenses}.

\noindent\textbf{Krum.} Krum~\citep{krum} tolerates $f$ Byzantine clients among $n$ clients by retaining only one local model that is the most likely to be benign as the global model. That is, Krum selects a single model as the global model in aggregation.
A generalization of Krum is $m$-Krum~\citep{krum} that selects $m$ client models with the $m$ lowest scores for aggregation, instead of choosing only one local model. This approach requires less than $\frac{n-m}{2}-1$ clients to be malicious. 

\begin{table*}[h]
    \centering
    \caption{Models and datasets for evaluations.}\label{tab:models_and_datasets}

    \begin{tabular}{lcccccc} 
    \toprule
       \textbf{Model} & ResNet20~\citep{he2016deep} & ResNet56~\citep{he2016deep} & CNN~\citep{mcmahan2017communication} & RNN (bi-LSTM)~\citep{mcmahan2017communication} & BERT~\citep{devlin2018bert} & Pythia-1B~\citep{biderman2023pythia} \\
       \midrule
       \textbf{Dataset} & CIFAR10~\citep{cifar} & CIFAR100~\citep{cifar} & FEMNIST~\citep{caldas2018leaf} & Shakespeare~\citep{shakespeare} & 20News~\citep{20news_dataset} & PubMedQA~\citep{QA_dataset} \\
       \bottomrule
    \end{tabular}
\end{table*}

\subsection{On-Aggregation Defenses}

On-aggregation defense functions modify the aggregation function to a robust version that tolerates or mitigates impacts of the potential adversarial client models. As an example, RFA (Robust Federated Aggregation)~\citep{rfa} computes a geometric median of the client models in each iteration as the aggregated model, instead of simply averaging the client models. RFA defense effectively mitigates the impact of poisoned client models, as the geometric median can represent the central tendency of the client models, and the median point is chosen in a way to minimize the sum of distances between that point and the other client models of the current FL iteration.
In practice, the geometric median is calculated using the Smoothed Weiszfeld Algorithm~\citep{rfa}.

\subsection{After-Aggregation Defenses}
After-aggregation defense functions modify the aggregation result, \textit{i}.\textit{e}., the global model, of each FL iteration to mitigate the effects of poisoned local models or protect the global model from potential adversaries. As an example, CRFL~\citep{xie2021crfl} clips the global model to bound the norm of the model each time after aggregation at the FL server. The FL server then adds Gaussian noise to the clipped global model before distributing the global model to the clients for the next FL iteration.


    

\subsection{Integration of a New Defense}\label{sec:new_defense}

To implement a self-designed defense mechanism, users should first determine the stages to inject the defense functions (\textit{i}.\textit{e}., before/on/after-aggregation), add a class for the new defense 
and implement the corresponding defense functions using the aforementioned APIs, \textit{i}.\textit{e}., $\mathit{defend}\_\mathit{before}\_\mathit{aggregation}(*)$, $\mathit{defend}\_\mathit{on}\_\mathit{aggregation}(*)$, and $\mathit{defend}\_\mathit{after}\_\mathit{aggregation}(*)$, to inject functions at appropriate stages of FL. Note that some defenses involve more than one stage; thus, users need to implement all relevant functions. Users should add the name of the defense to the enabler functions 
to activate the injected function at the different stages of FL. 
The approach computes some scores using local models submitted by clients, and uses the scores to identify outlier local models before aggregating the local models. As such process only happens before aggregation, we only need to implement $\mathit{defend}\_\mathit{before}\_\mathit{aggregation}(*)$ for the defense class, and include the name of the defense in $\mathit{is}\_\mathit{defense}\_\mathit{after}\_\mathit{aggregation}()$.

\section{Evaluations}\label{sec:exp}

This section presents a comprehensive evaluation of FedSecurity to benchmark some well-known attacks and defenses in FL. 

\begin{figure*}[htbp]
    \centering
    \begin{minipage}{0.24\textwidth}
        \centering
        \includegraphics[width=\textwidth]{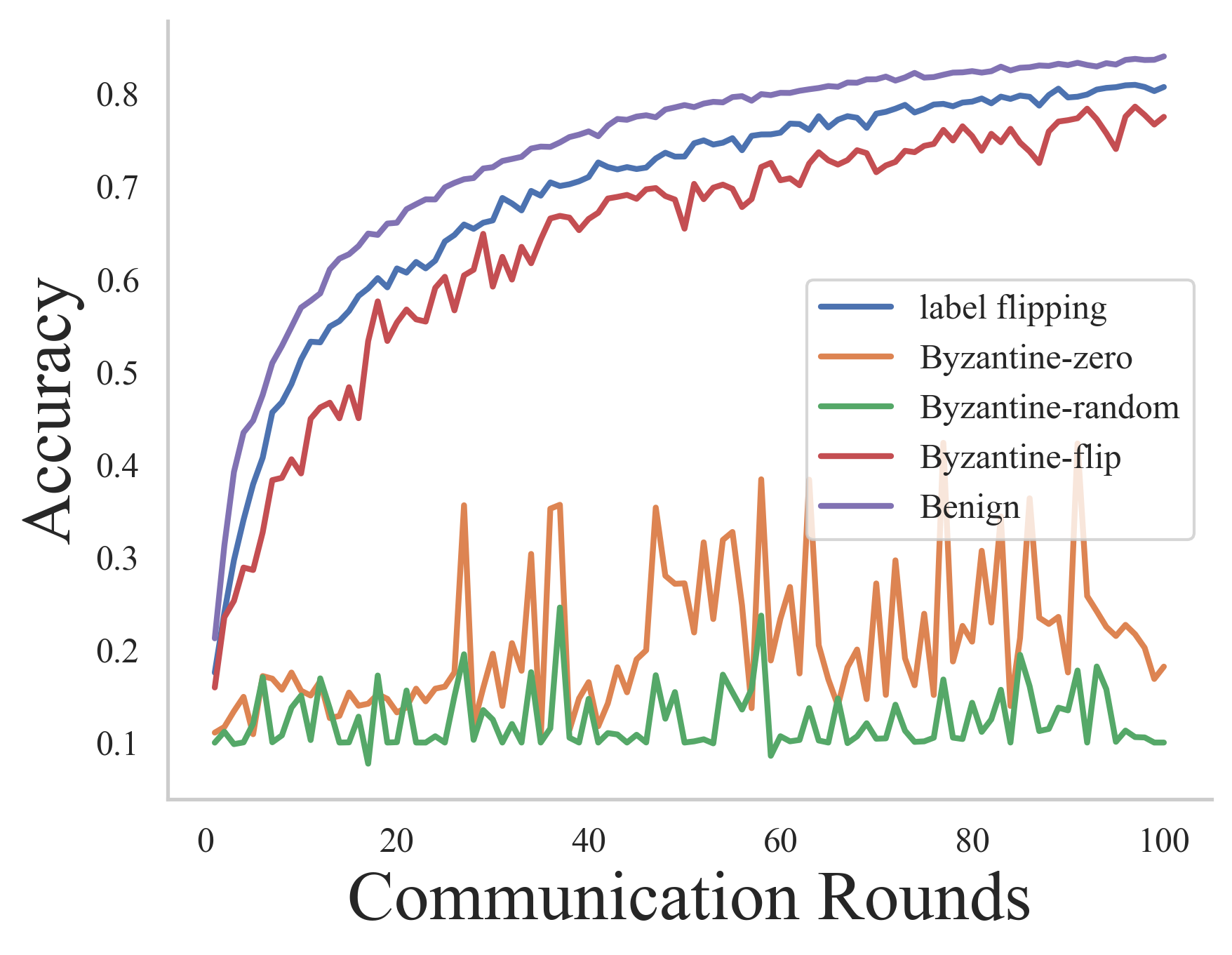}
        \caption{Attack comparison. }
        \label{fig:attacks_comparison_all}
    \end{minipage}%
    \begin{minipage}{0.24\textwidth}
        \centering
        \includegraphics[width=\textwidth]{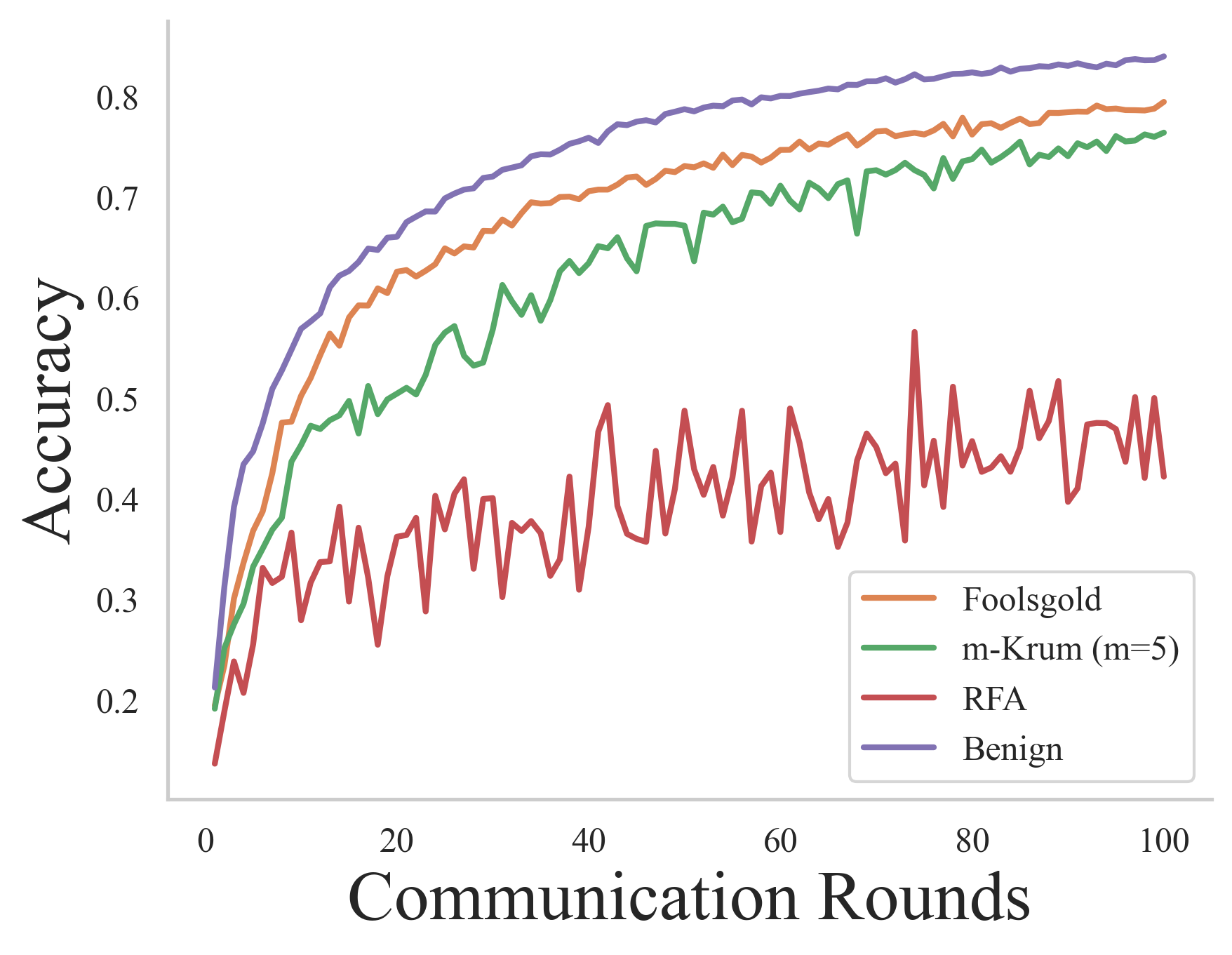}
        \caption{Defense comparison.}
        \label{fig:compare_defenses}
    \end{minipage}%
    \hfill
    \centering
    \begin{minipage}{0.24\textwidth}
        \centering
        \includegraphics[width=\linewidth]{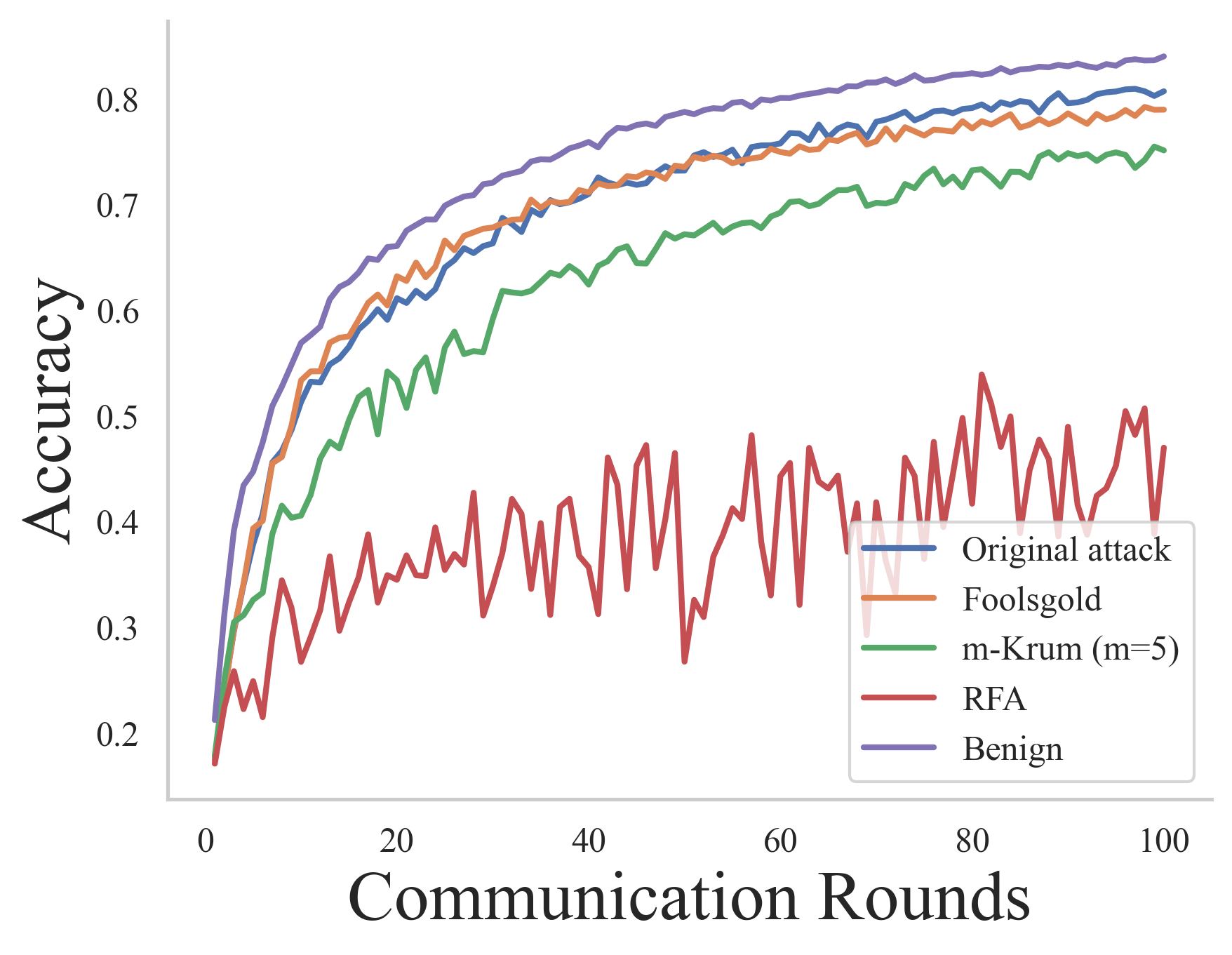}
        \caption{Label flipping exps.}
        \label{fig:label_flipping_with_defenses}
    \end{minipage}%
    \begin{minipage}{0.24\textwidth}
	\centering	\includegraphics[width=.99\linewidth]{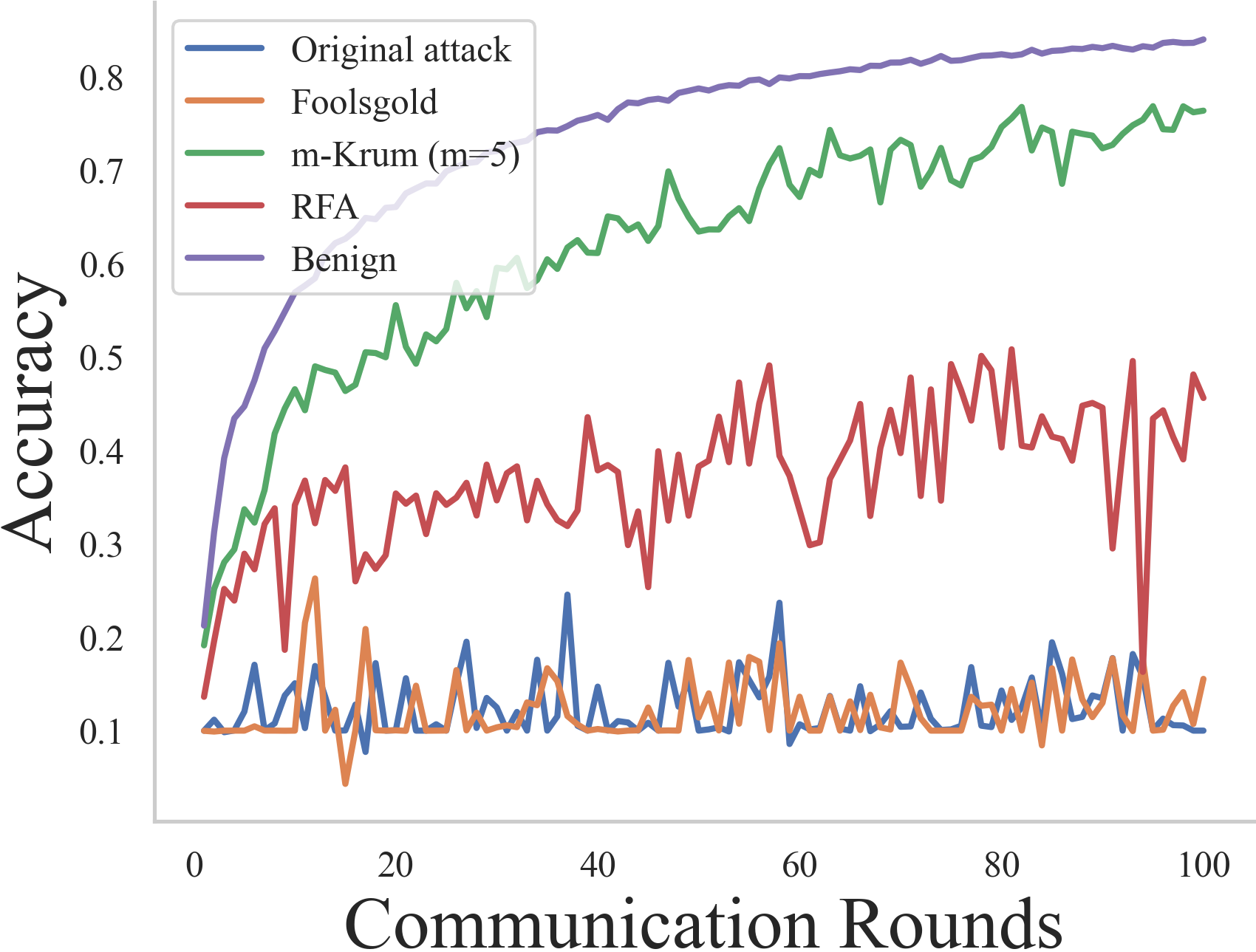}
	\caption{Random-Byzantine exps.}\label{fig:byzantine_random_with_defenses}
   \end{minipage}%
\end{figure*}

\begin{figure*}[htbp]
    \centering
    \begin{minipage}{0.24\textwidth}
        \centering
        \includegraphics[width=.99\linewidth]{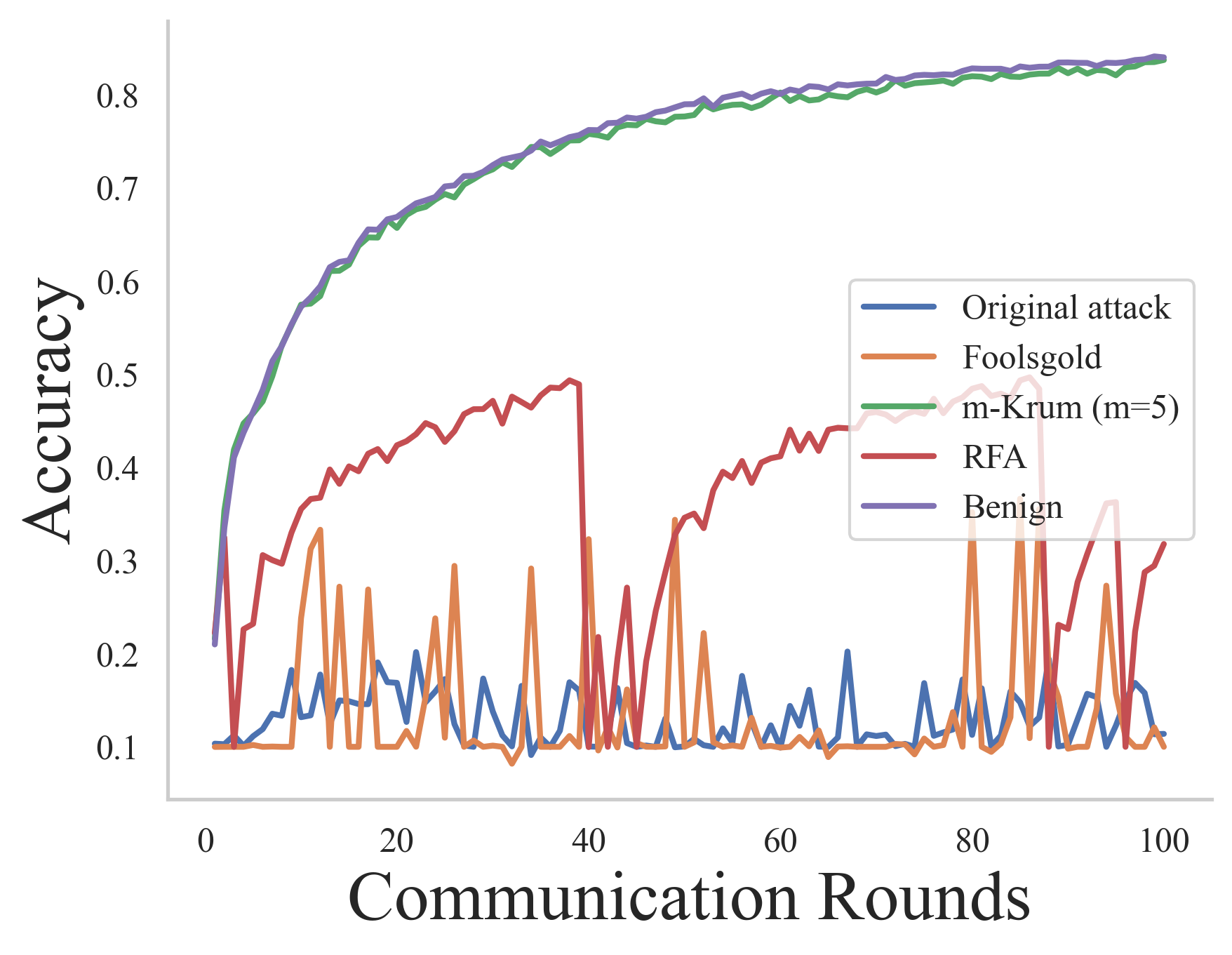}
   \caption{I.I.D.~data evaluations.}\label{fig:iid_exp}
    \end{minipage}%
    \begin{minipage}{0.24\textwidth}
        \centering
        \includegraphics[width=.99\linewidth]{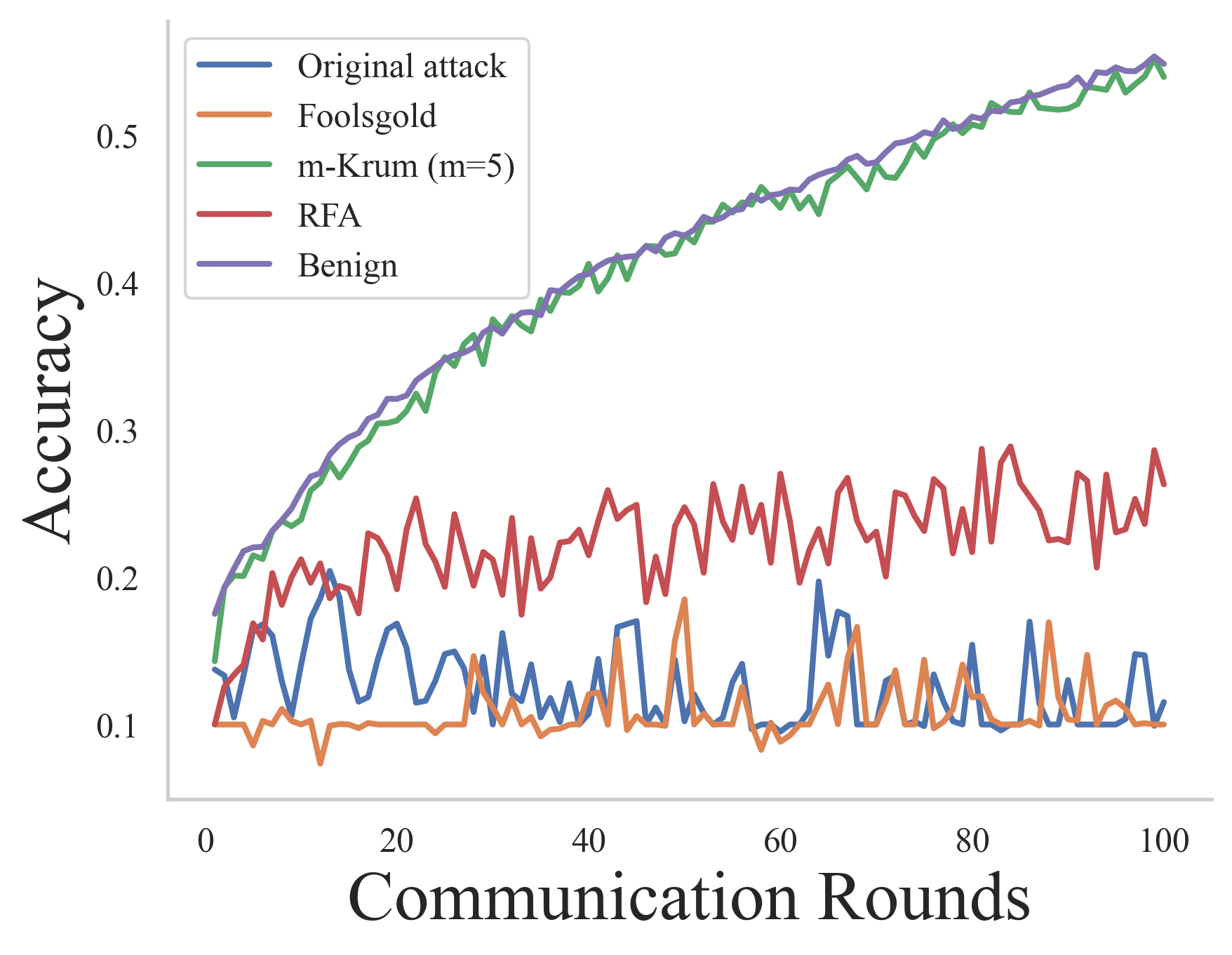}
   \caption{Scale \# clients to 100.} \label{fig:100clients_exp}
    \end{minipage}%
    \hfill
    \centering
    \begin{minipage}{0.24\textwidth}
        \centering\includegraphics[width=.99\linewidth]{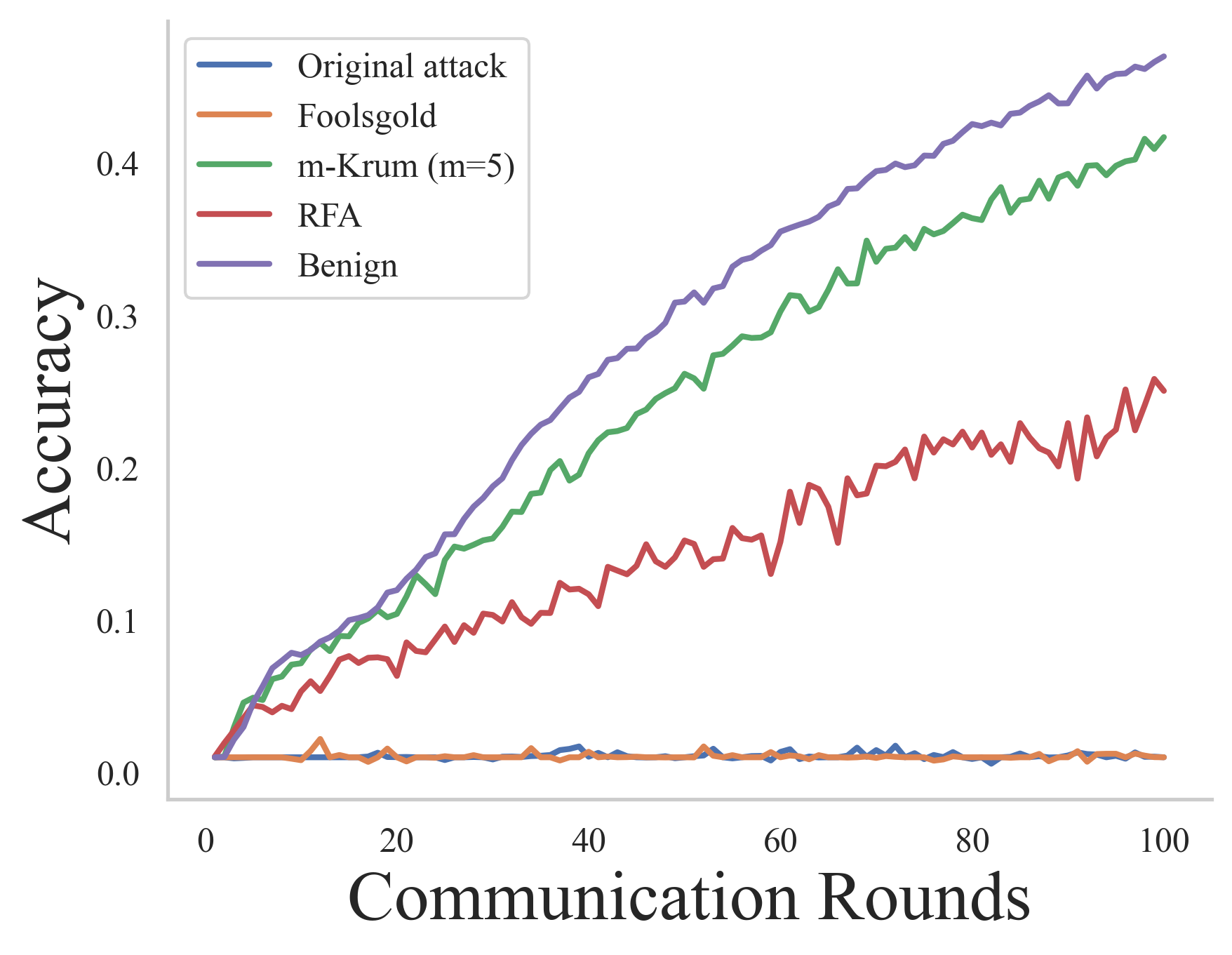}
        \caption{ResNet56 (CV). }
        \label{fig:resnet56_exp}
    \end{minipage}%
    \begin{minipage}{0.24\textwidth}
	\centering	\includegraphics[width=.99\linewidth]{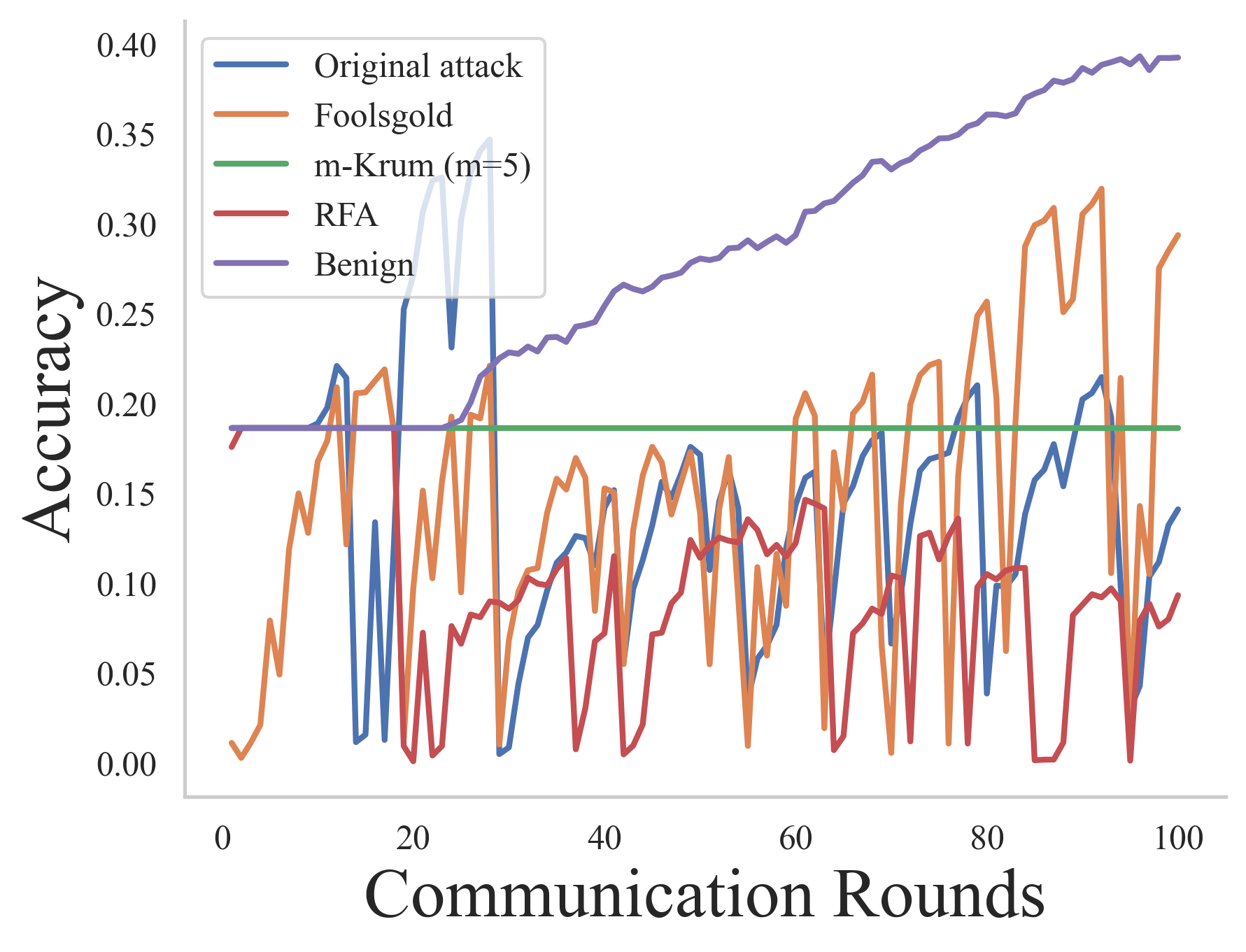}
        \caption{RNN (NLP). }
        \label{fig:nlp_exp}
   \end{minipage}%
\end{figure*}

\begin{figure}[htbp]    
    \begin{minipage}{0.24\textwidth}
        \centering
        \includegraphics[width=\textwidth]{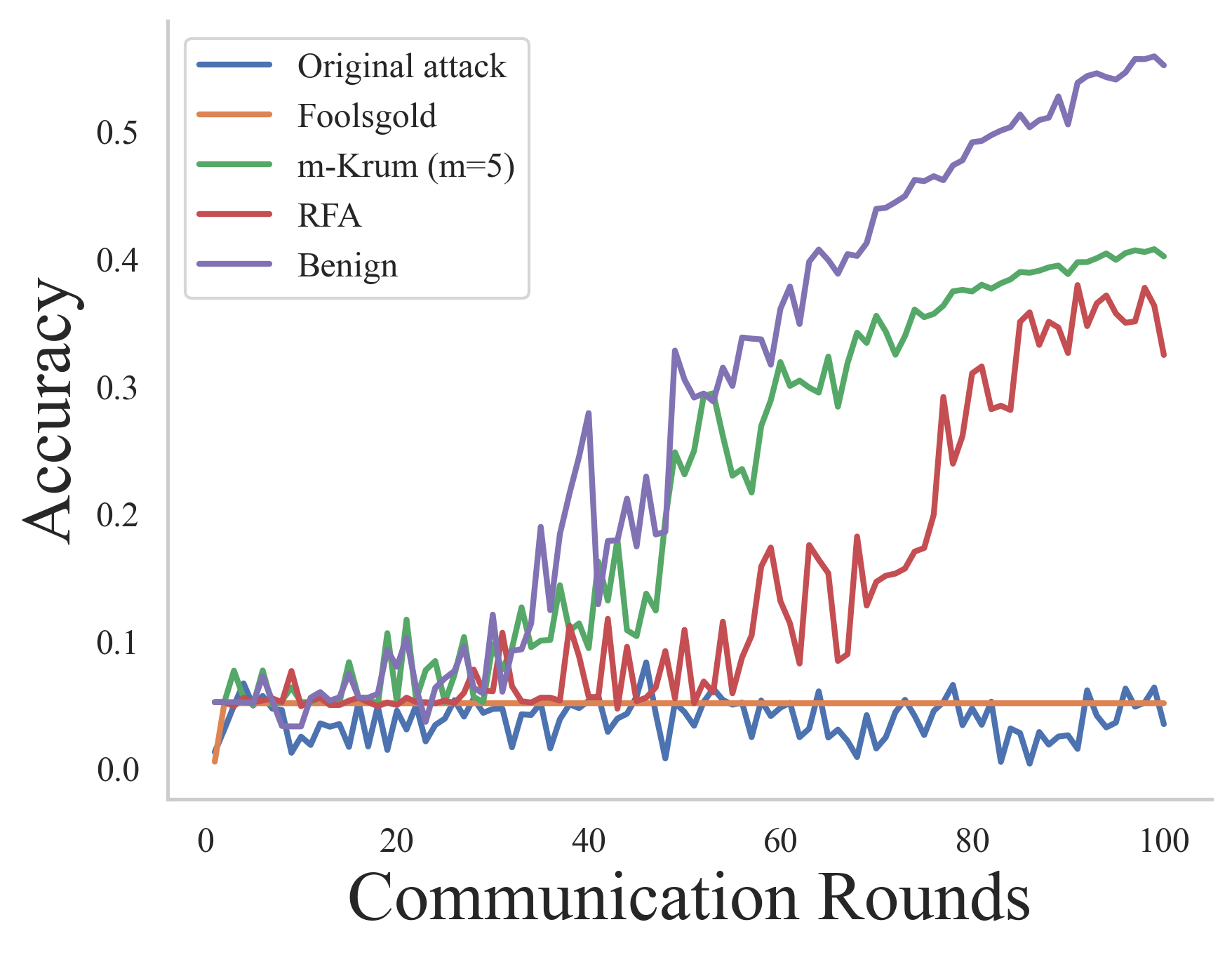}
        \caption{CNN (CV). }
        \label{fig:cnn_exp}
    \end{minipage}%
    \centering
    \begin{minipage}{0.24\textwidth}
        \centering
        \includegraphics[width=\textwidth]{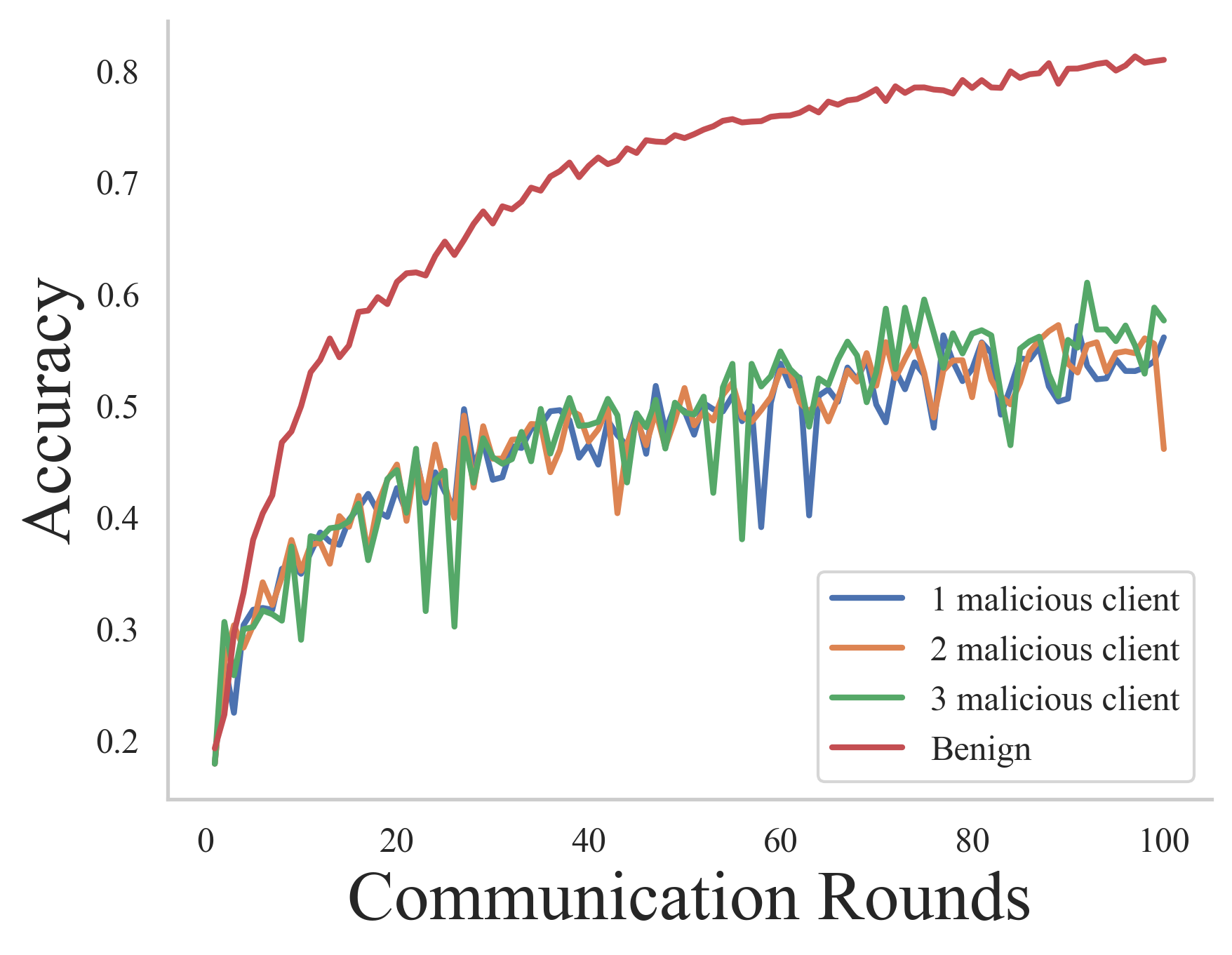}
        \caption{Varying \# adversaries. }
        \label{fig:varying_malicious_client_num}
    \end{minipage}%

\end{figure}

\noindent\textbf{Experimental setting. } 
A summary of datasets and models for evaluations can be found in Table~\ref{tab:models_and_datasets}. We utilize FedAVG in our experiments.  By default, we employ ResNet20 and the non-i.i.d.~CIFAR10 dataset (partition parameter $\alpha = 0.5$), as the non-i.i.d.~setting to capture real-world scenarios closely. We further extend our evaluations to i.i.d.~cases and various other models and datasets.
For evaluations on LLMs, we utilize FedLLM~\citep{fedllm} that trains LLMs in a federated manner. We employ the Pythia-1B model~\citep{biderman2023pythia} and PubMedQA~\citep{jin2019pubmedqa}, a non-i.i.d. biomedical research dataset that contains 212,269 questions for question answering. We utilize the ``artificial'' subset for training and the ``labelled'' subset for testing.
Evaluations are conducted on a server with 8 NVIDIA A100-SXM4-80GB GPUs.  

By default, we use 10 clients for FL training, corresponding to real-world FL applications where the number of clients is typically less than 10, especially in business-to-business (B2B) scenarios. We also increase the number of clients to 100 in \textbf{Exp 5}, and set the number of clients to
70 in the real-world experiment, where we utilize real edge devices from Theta network~\citep{theta_network} to showcase the scalability of FedSecurity (\textbf{Exp 10}). 
Unless otherwise noted, we set the percentage of malicious clients to 10\%, and evaluate results with the accuracy of the global model. We employ three attack mechanisms, including label flipping~\citep{label_flipping} and Byzantine attacks of random mode and flipping mode~\citep{chen2017distributed,fang2020local,xu2022byzantine}. For the label flipping attack, we set the attack to modify the local and test data labels of malicious clients from label 3 to label 9 and label 2 to label 1. We utilize three defense mechanisms: $m$-Krum~\citep{krum}, Foolsgold~\citep{foolsgold}, and RFA~\citep{rfa}. For $m$-Krum, we set $m$ to 5, which means 5 out of 10 local models participate in aggregation in each training round.

\subsection{Evaluations on FL}

\noindent\textbf{Exp 1: Attack Comparisons.} 
We evaluate the impact of attacks on test accuracy, using a no-attack scenario as a baseline. As illustrated in Figure~\ref{fig:attacks_comparison_all}, Byzantine attacks, specifically in the random and zero modes, substantially degrade accuracy. In contrast, the label flipping attack and the flipping mode of the Byzantine attack show a milder impact on accuracy. This can be attributed to the nature of Byzantine attacks, where Byzantine attackers would prevent the global model from converging, especially for the random mode that generates weights for models arbitrarily, causing the most significant deviation from the benign local model. In subsequent experiments, unless specified otherwise, we employ the Byzantine attack in the random mode as the default attack, as it provides the strongest impact compared with the other three attacks.

\noindent\textbf{Exp 2: Defense Comparisons.} 
We investigate potential impact of defense mechanisms on accuracy in the absence of attacks, \textit{i}.\textit{e}., whether defense mechanisms inadvertently degrade accuracy when all clients are benign. We incorporate a scenario without any defense or attack as our baseline. 
As illustrated in Figure~\ref{fig:compare_defenses}, it becomes evident that when all clients are benign, involving defense strategies to FL training might lead to a reduction in accuracy. This decrease might arise from several factors: the exclusion of some benign local models from aggregation, \textit{e}.\textit{g}., as in $m$-Krum, adjustments to the aggregation function, \textit{e}.\textit{g}., as in RFA, or re-weighting local models, \textit{e}.\textit{g}., as in Foolsgold. 
Specifically, the RFA defense mechanism significantly impacts accuracy as it computes a geometric median of the local models instead of leveraging the original FedAVG optimizer, which introduces a degradation in accuracy.



\begin{figure*}[htbp]
    
    \begin{minipage}{0.33\textwidth}
        \centering
        \includegraphics[width=\textwidth]{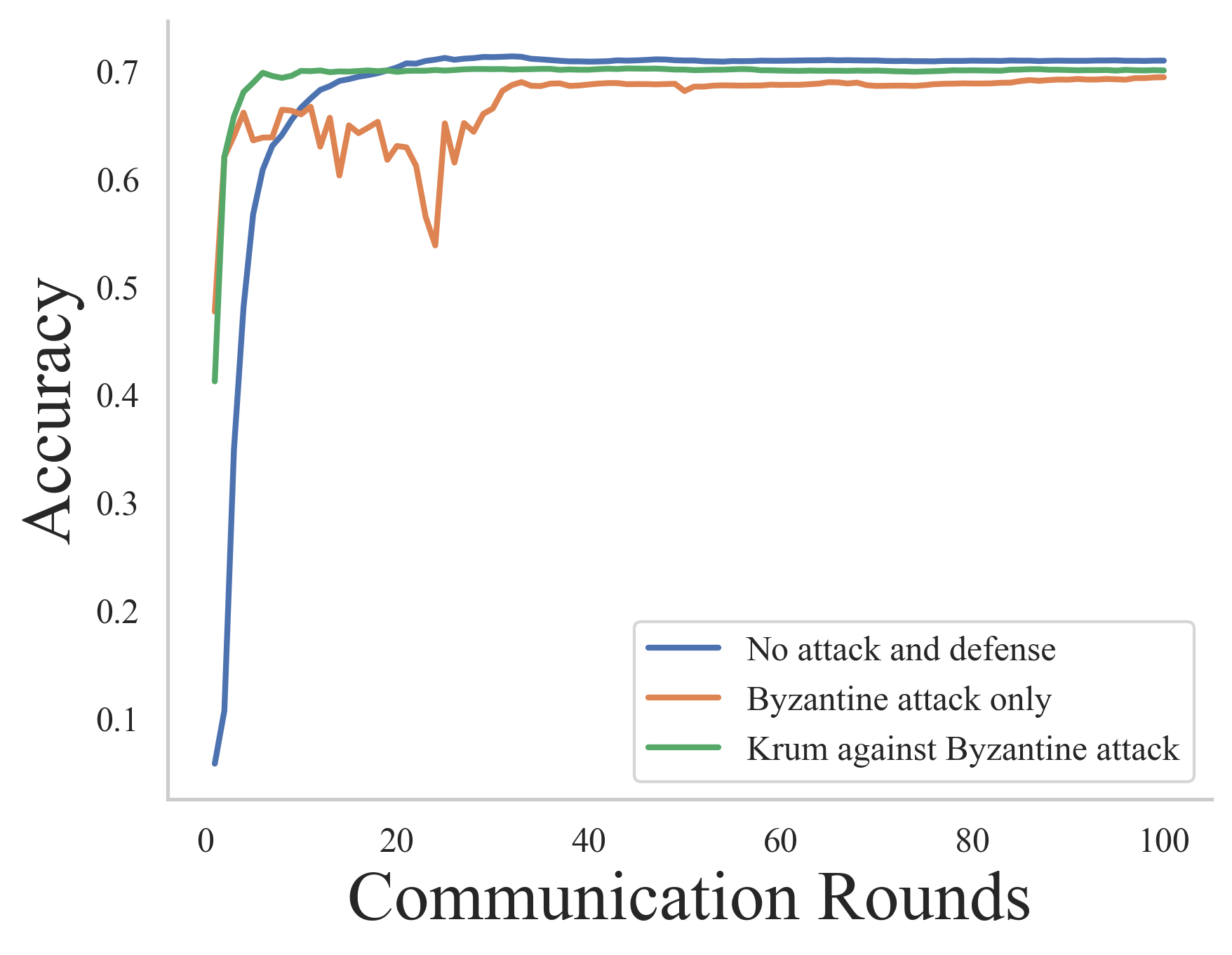}

        \caption{BERT evaluations.}
        \label{fig:bert_exp}
    \end{minipage}%
    \begin{minipage}{0.33\textwidth}
        \centering
        \includegraphics[width=\textwidth]{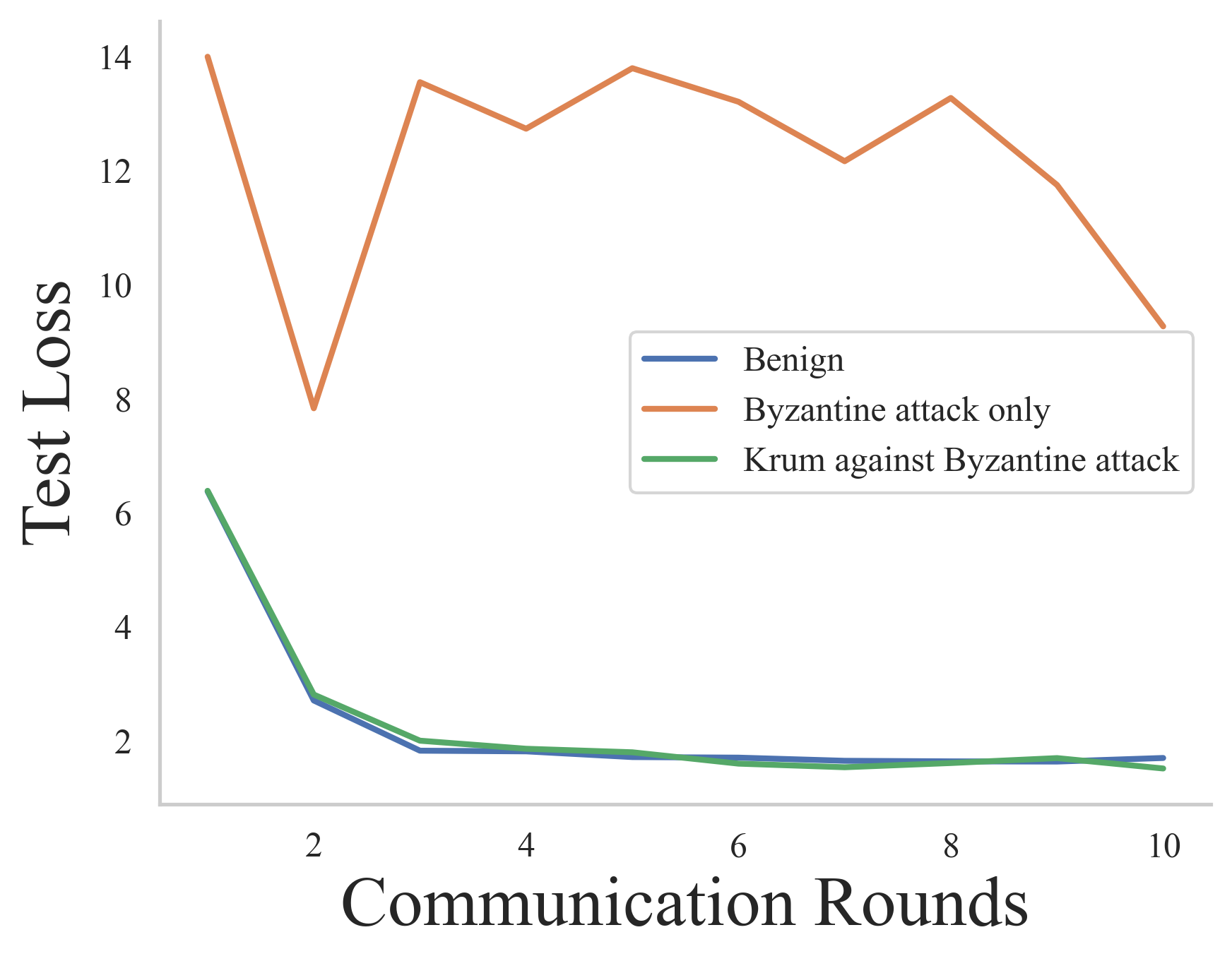}

        \caption{Pythia-1B evaluations.}
        \label{fig:llm_exp}
    \end{minipage}%
    \begin{minipage}{0.33\textwidth}
        \centering
        \includegraphics[width=\textwidth]{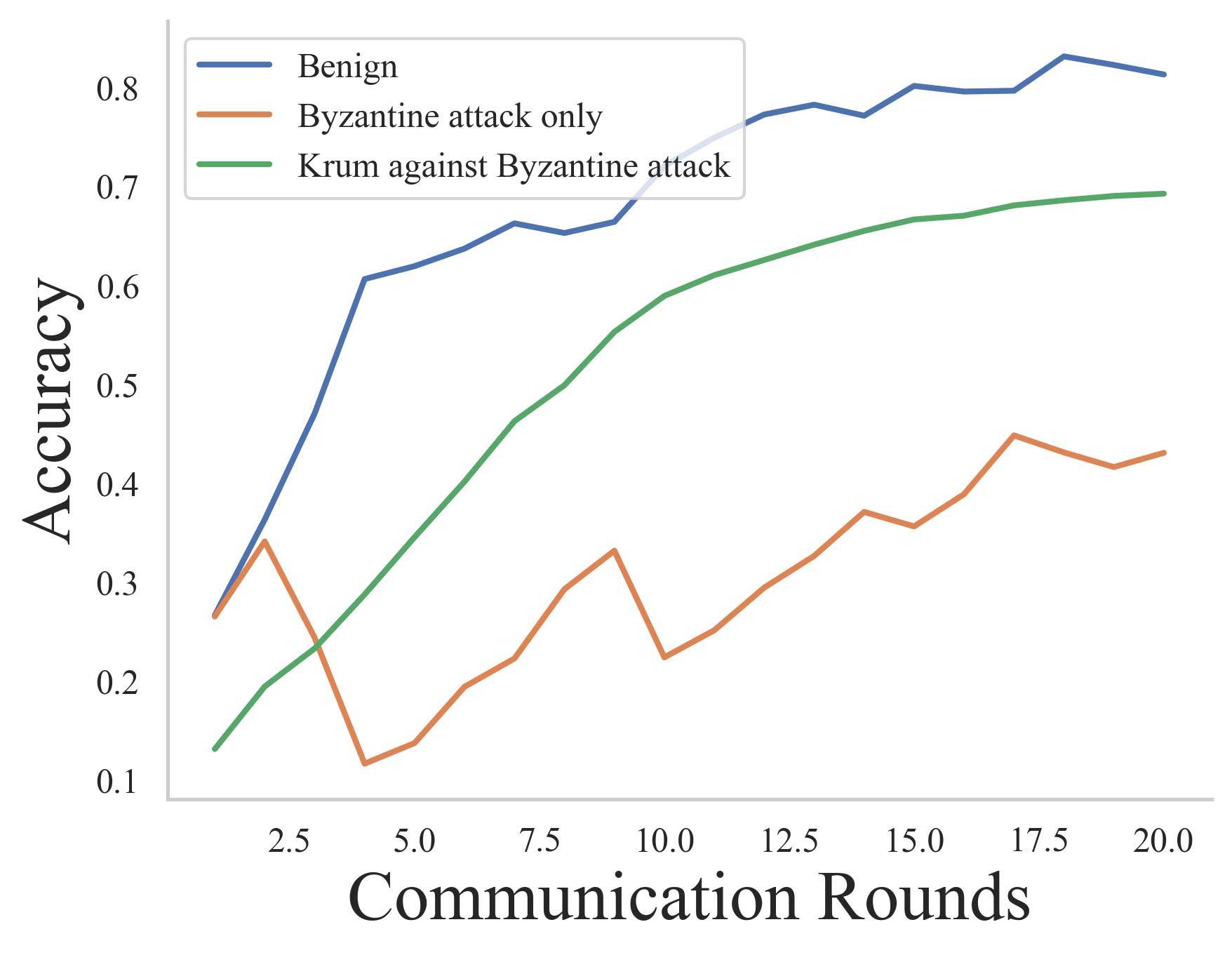}

    \caption{Real-world application evaluation.
    }
    \label{fig:real_world simulation_krum}
    \end{minipage}%

\end{figure*}


\noindent\textbf{Exp 3: Evaluations of defense mechanisms against activated attacks.} This experiment evaluates the effect of defense mechanisms against some attacks. We include two baseline scenarios: 1) an ``original attack'' scenario with an activated attack without any defense in place, and 2) a ``benign'' scenario with no activated attack or defense. 
We select label flipping attack and the random mode of Byzantine attack based on their impacts in \textbf{Exp1}, where label flipping has the least impact and the random mode of Byzantine attack exhibits the largest impact, as shown in Figure~\ref{fig:attacks_comparison_all}. 
Results for the label flipping and the random mode of   Byzantine attacks are in Figure~\ref{fig:label_flipping_with_defenses} and Figure~\ref{fig:byzantine_random_with_defenses}, respectively. 
These results indicate that the defenses may contribute to minor improvements in accuracy for low-impact attacks, \textit{e}.\textit{g}., Foolsgold in Figure~\ref{fig:label_flipping_with_defenses}.  In certain cases, it is noteworthy that the defensive mechanisms may inadvertently compromise accuracy, such as the case with RFA in Figure~\ref{fig:label_flipping_with_defenses}.
For high-impact attacks, such as the Byzantine attack of the random mode, Krum exhibits resilience, effectively neutralizing the negative impact of the attacks, as shown in Figure~\ref{fig:byzantine_random_with_defenses}. 

\noindent\textbf{Exp 4: Evaluations on i.i.d.~data.}
We select the random mode of the Byzantine attack, 
and employ Foolsgold, $m$-Krum ($m=5$), and RFA to counteract the adverse effects of this attack. As shown in Figure~\ref{fig:iid_exp}, $m$-Krum is the most effective one among all the defense mechanisms, where the test accuracy is close to the case where all the FL clients are honest, \textit{i}.\textit{e}., no attack scenario.

\noindent\textbf{Exp 5: Scaling the number of clients to 100.}
We scale the number of clients to 100 and evaluates the defense mechanisms against the random mode of the Byzantine attack. We employ Foolsgold, $m$-Krum (with $m=5$), and RFA to counteract the adverse effects of this attack. As shown in Figure~\ref{fig:100clients_exp}, $m$-Krum is the most effective one among all the defense mechanisms, and the test accuracy is very close to the case where no attack happens. That is because in each FL iteration, 
$m$-Krum selects 5 local models that are more likely to be benign, which can represent the other local models, thus can achieve comparable accuracy compared with the benign case. 

\noindent\textbf{Exp 6: Evaluations on different models.}
We evaluate defense mechanisms against the random mode of the Byzantine attack with different models and datasets, including: \textit{i}) ResNet56 + CIFAR100, \textit{ii}) RNN + Shakespeare, and \textit{iii}) CNN + FEMNIST. The results are shown in Figures~\ref{fig:resnet56_exp},~\ref{fig:nlp_exp}, and~\ref{fig:cnn_exp}, respectively. The results show that while the defense mechanisms can mitigate the impact of attacks in most cases,
some attacks may fail some tasks, \textit{e}.\textit{g}., $m$-Krum fails RNN in Figure~\ref{fig:nlp_exp}, and Foolsgold fails CNN in Figure~\ref{fig:cnn_exp}. This is because the two defense mechanisms either select several local models for aggregation in each FL training round, or significantly re-weight the local models, which may eliminate some local models that are important to the aggregation in the first several FL training iterations, leading to unchanged test accuracy in later FL iterations.

\noindent\textbf{Exp 7: Varying the number of malicious clients.}
This experiment evaluates the impact of varying numbers of malicious clients on test accuracy. We utilize $m$-Krum to protect against 1, 2, and 3 malicious clients out of 10 clients in each FL training round. 
As shown in Figure~\ref{fig:varying_malicious_client_num}, 
the test accuracy remains relatively consistent across different numbers of malicious clients, as in each FL training round, $m$-Krum selects a local model that is the most likely to be benign to represent the other models, effectively minimizing the impact of malicious client models on the aggregation.

\begin{figure*}
    \centering
    \includegraphics[width=17cm]{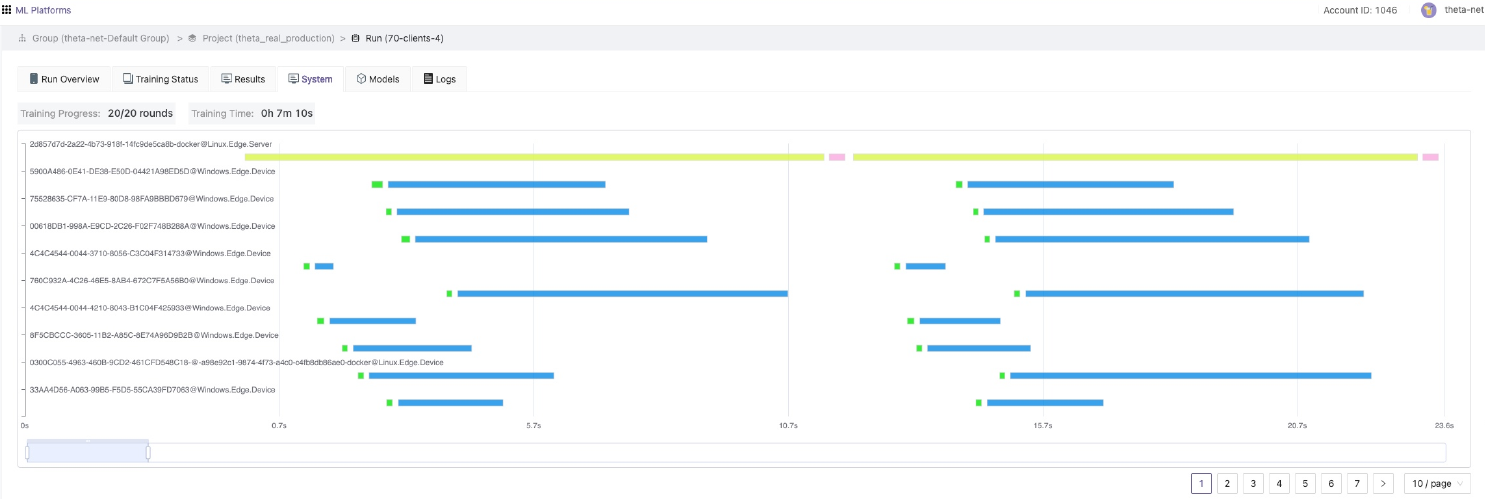}
    \caption{Real-world application. Yellow: aggregation server waiting time; pink: aggregation time; green: client training time; blue: client communication.
    }
    \label{fig:real_world simulation}
\end{figure*}

\begin{figure*}
    \centering
    \includegraphics[width=17cm]{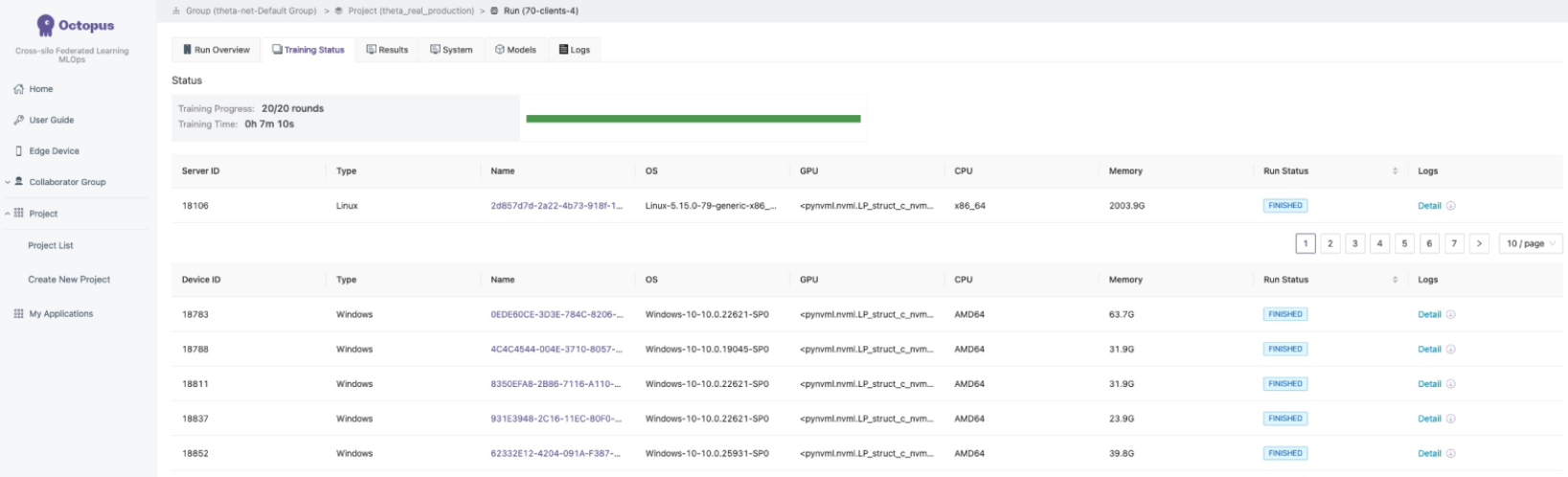}
    \caption{Real-world application: training status of devices. 
    }
    \label{fig:real_world_training}
\end{figure*} 
 
\subsection{Evaluations on Federated LLMs}\label{sec:llm}

We employ two LLMs, BERT~\citep{devlin2018bert} and Pythia~\citep{biderman2023pythia}, to showcase the scalability of FedSecurity and its applicability to federated LLM scenarios. We notice that some defenses (\textit{e}.\textit{g}., Foolsgold~\citep{foolsgold}) that require memorizing intermediate results, such as models of previous FL training rounds, might encounter limitations when integrated with LLMs due to the significant cache introduced. Considering this, we utilize $m$-Krum for our experiments, as it does not require storing intermediate results and demonstrates consistent performance in most of our previous experiments.

\noindent\textbf{Exp 8: Evaluations of Krum against model replacement backdoor attack on BERT. } This experiment utilizes BERT~\citep{devlin2018bert} and the 20 news dataset~\citep{20news_dataset} for a classification task. We employ 10 clients and set 1 client to be malicious in each FL training round. We set $m$ to 5 in $m$-Krum, \textit{i}.\textit{e}., 5 out of 10 local models participate in aggregation in each FL training round. Results in Figure~\ref{fig:bert_exp} show that $m$-Krum effectively mitigates the adversarial effect, bringing the accuracy closer to the level of the attack-free case.

\noindent\textbf{Exp 9: Evaluations of Krum against the Byzantine attack on Pythia-1B. } 
We employ 7 clients for FL training, and 1 out of 7 clients is malicious in each round of FL training. We set the $m$ parameter in $m$-Krum to 2, signifying that 2 out of 7 submitted local models participate in the aggregation in each FL training round. The performance is evaluated with the test loss. Results in Figure~\ref{fig:llm_exp} show that Byzantine attack significantly increases the test loss during training. Nevertheless, $m$-Krum effectively mitigates the adversarial effect.

\subsection{Evaluation in Real-World Applications}\label{sec: real_world}
To demonstrate the scalability of our benchmark, we include an experiment using real-world devices, instead of simulations.

\noindent\textbf{Exp10: Evaluations in real-world applications.} 
We utilize edge devices from the Theta network~\citep{theta_network} to validate the scalability of FedSecurity to real-world applications. The FL client package is integrated into Theta's edge nodes, which periodically fetches data from the Theta back-end. Subsequently, the FL training platform capitalizes on these Theta edge nodes and their associated data to train, fine-tune, and deploy machine learning models.

We select $m$-Krum as the defense and the Byzantine attack of random mode as the attack. Considering the challenges posed by real-world environments, such as devices equipped solely with CPUs (lacking GPUs), potential device connectivity issues, network latency, and limited storage on edge devices (for instance, some mobile devices might have less than 500MB of available storage), we choose a simple task by employing the MNIST dataset for a logistic regression task.
In our experimental setup, we deploy 70 client edge devices, designating 7 of these as malicious for each FL training round. For $m$-Krum, we set $m$ to 35, meaning that 35 out of the 70 local models are involved in aggregation during each FL training round. As illustrated in Figure~\ref{fig:real_world simulation_krum}, $m$-Krum mitigates the adversarial effect of the random-mode Byzantine attack. 
We also include a screenshot of the platform, as shown in Figure~\ref{fig:real_world simulation} for the FL training process and Figure~\ref{fig:real_world_training} for the training status of each real-world device.


\section{Conclusion}


This paper presents FedSecurity, a benchmark 
for adversarial attacks and corresponding defense strategies in FL. FedSecurity contains two components: FedAttacker that simulates various attacks that can be injected during FL training, and FedDefender that facilitates defense strategies to mitigate the impacts of these attacks. 
\new{The limitation of FedSecurity is that it does not support asynchronous FL and vertical FL yet.}
FedSecurity is open-sourced, and we welcome contributions from the research community to enrich the benchmark repository with novel attack and defense strategies to foster a diverse, comprehensive, and robust foundation for ongoing research in FL security.


\bibliographystyle{ACM-Reference-Format}
\balance
\bibliography{reference}

\appendix


\end{document}